\documentclass[11pt]{article}

\usepackage[usenames,dvipsnames,table]{xcolor}

\pdfoutput=1 % if your are submitting a pdflatex (i.e. if you have
\usepackage{jheppub} % for details on the use of the package, please
\usepackage{graphicx}
\usepackage{amsmath, amssymb}
\usepackage{youngtab}
\usepackage{manfnt}
\usepackage{float}
\usepackage{placeins}
\usepackage{ytableau}
\ytableausetup{boxsize=.6em,centertableaux}
\usepackage{arydshln}
\usepackage{dsfont}
\usepackage{slashed}
\usepackage{subcaption}
\usepackage{comment}

\usepackage{tikz}
\usetikzlibrary{positioning}

\def\be{\begin{eqnarray}}
\def\ee{\end{eqnarray}}
\newcommand{\nn}{\nonumber}

\def\Dslash{\,\,{\raise.15ex\hbox{/}\mkern-12mu D}}
\def\Dbarslash{\,\,{\raise.15ex\hbox{/}\mkern-12mu {\bar D}}}
\def\delslash{\,\,{\raise.15ex\hbox{/}\mkern-9mu \partial}}
\def\delbarslash{\,\,{\raise.15ex\hbox{/}\mkern-9mu {\bar\partial}}}
\def\pslash{\,\,{\raise.15ex\hbox{/}\mkern-9mu p}}
\def\calDslash{\,\,{\raise.15ex\hbox{/}\mkern-12mu {\cal D}}}

\newcommand{\sym}{\ydiagram{2}}
\newcommand{\anti}{\ydiagram{1,1}}

\newcommand{\trho}{\tilde{\rho}}
\newcommand{\tchi}{\tilde{\chi}}
\newcommand{\tilb}{\tilde{\mathcal{B}}}
\newcommand{\bJ}{\bar{J}}
\newcommand{\bO}{\bar{\mathcal{O}}}

\title{RG Flows in 2d Chiral Gauge Theories}

\author{Kaan \"{O}nder}

\affiliation{Department of Applied Mathematics and Theoretical Physics \\ University of Cambridge, CB3 0WA, UK}

\emailAdd{ko354@cam.ac.uk}

\abstract{We study the dynamics of two-dimensional chiral $SU(N)$ gauge theories with fermions in the symmetric, anti-symmetric and fundamental representations. A consistent infra-red limit of these theories consists of certain coset CFTs. There is also a free fermion phase which shares the same central charge and 't Hooft anomalies but does not coincide with the coset models. We show that these two theories sit on a conformal manifold of infra-red theories and are related by a current-current deformation. We further consider extensions of these theories by adding Dirac fermions and comment on possible RG flows.}

\begin{document}

\maketitle
\flushbottom
%%%%%%%%%%%%%%%%%%%%%   NOTES START  %%%%%%%%%%%%%%%%%%%%%%%%%%%
%%%%%%%%%%%%%%%%%%%%%%%%%%%%%%%%%%%%%%%%%%%%%%%%%%%%%%%%%%%%%%%%%

\section{Introduction}
A central challenge of any strongly coupled gauge theory is determining the low energy dynamics. This is an inherently non-perturbative problem and, in general, can not be answered rigorously. Two dimensional gauge theories with massless quarks are examples of such theories. Unlike in four-dimensions, where the theory becomes strongly coupled by virtue of the beta-function and hence depends on the number of quarks, these two dimensional theories are strongly coupled for any number of quarks. This is a consequence of the Yang-Mills coupling having a positive mass dimension. 

The 2d dynamics differ from their 4d counterparts, notably as the Coleman-Mermin-Wagner theorem rules out spontaneous breaking of continuous global symmetries \cite{mermin,coleman}. There are however similarities such as confinement, discrete chiral symmetry breaking and dynamical mass generation. As the study of these phenomena are more tractable in 2d, this makes the 2d models a good testing ground for understanding these non-pertubative properties. Together with condensed matter applications, this has led to a long list of studies of these theories over the years (see \cite{Schwinger,tHooftlargeN,Wittenbaryon,Kutasov,Donahue,Komargodski,Cherman,Dempsey,Dempsey2,Baluni,Durgut} for a far from comprehensive list).

The dynamics of chiral gauge theories  is particularly interesting. They can behave very differently from their vector-like counterparts, not least because they evade the Vafa-Witten-Weingarten theorems \cite{vw,wein}. Furthermore, the Nielsen-Ninomiya theorem \cite{NN1,NN2} presents a challenge in lattice regularising these theories, ruling out any numerical analysis (for recent developments on putting 2d chiral fermions on the lattice, while preserving Abelian symmetries, see \cite{wang,Wang1,Zeng,Wang2}). Understanding the dynamics of chiral gauge theories rigorously remains one of the big open questions in QFT.

In interesting recent work, Delmastro, Gomis and Yu \cite{Delmastro,Delmastro2} study the dynamics of general 2d gauge theories coupled to massless Weyl fermions. They consider gauge theories with a compact, connected gauge group $G$ coupled to left and right-handed Weyl fermions in representations $R_l$ and $R_r$ of $G$ respectively. They propose that the low-energy dynamics of these theories are captured by the coset CFT/TQFT
\be
\mathcal{T}_{\text{IR}}=\frac{U\left(\operatorname{dim} R_{\ell}\right)_{1}}{G_{\mu\left(R_{\ell}\right)}} \times \frac{U\left(\operatorname{dim} R_{r}\right)_{1}}{G_{\mu\left(R_{r}\right)}}
\label{QCDcoset1}\ee
where $\mu(R)$ is the Dynkin index of the representation $R$. This RG flow was also previously proposed by Komargodski et al. \cite{Komargodski} in the context of adjoint QCD.

The form of this low-energy coset has a very specific implication. Any chiral flavour symmetry of the deep UV theory is preserved along the RG flow triggered by the gauge coupling. In other words, any holomorphic current of the UV theory remains holomorphic along the RG flow and descends to a holomorphic current in the deep IR CFT. The full chiral algebras are preserved along the flow. The same statements hold true for anti-holomorphic currents. Specifically, $U(1)$ charges of left and right-moving operators can not mix under RG flow.

This is a very particular property of the coset proposal \eqref{QCDcoset1} and will not hold generically. Along general RG flows, conformal symmetry is violated and the charges need not factorise to left and right movers. When we finally hit the infra-red CFT, the IR currents will be some combination of the UV currents and will not retain the same holomorphic factorisation as the UV. Said in another way, the levels of the Abelian UV and IR chiral algebras will generically not match. 

The flow to the coset \eqref{QCDcoset1} and consequently the conservation of holomorphic factorisation along RG flows only hold for UV theories composed of Weyl fermions interacting via gauge interactions. This conservation is no longer the case if we consider deformations of the UV theory such as four-fermion interactions or if we include scalar fields interacting via Yukawa terms. Small deformations of the UV theory map to small perturbations of the coset \eqref{QCDcoset1} and can be systematically studied. Conversely, large UV deformations can lead to completely different RG flows and are much more difficult to analyse.

In this paper, we study the effects of such deformations on certain 2d non-Abelian chiral gauge theories. These theories have gauge group $SU(N)$ and contain a right-moving Weyl fermion in either the
symmetric or anti-symmetric representation, together with a collection of further left-moving fermions in the anti-fundamental. Applying the coset proposal \eqref{QCDcoset1} to these theories with only gauge interactions gives a candidate low-energy limit. There is also a different, rather natural candidate IR phase which simply consists of free, massless fermions which saturate all anomalies. This was proposed by Tong \cite{DTsm} in the context of symmetric mass generation. These free fermions have mixed global $U(1)$ charges and, unlike the coset model, do not have the same holomorphic factorisation as the UV chiral theories.

How are these coset and free fermion phases related to one another? We study this question and provide evidence that four-fermion and Yukawa deformations of the UV gauge theories change the RG flow and parameterise a conformal manifold of low energy theories. These deformations continuously connect the coset and free fermion descriptions which sit at different ends of this conformal manifold. The mixing of $U(1)$ charges under RG flow plays an important role in this connection, which we analyse in detail.

We start in Section \ref{p0theories} by introducing the chiral gauge theories under consideration. We present the candidate coset and free fermion phases, discuss their deformations, show that they can be continuously connected and conjecture that they live on a conformal manifold. In Section \ref{general}, we consider extending these chiral theories by adding further anti-fundamental Dirac fermions. These theories are 2d analogues of the well known 4d  Bars-Yankielowicz models \cite{by}. This time, the coset and free fermion phases have differing central charges and are not connected by a conformal manifold. We show that the free fermion phase may be an unstable fixed point and comment on possible RG flows. Finally, we discuss discrete symmetries in Appendix \ref{discrete} and analyse the one-loop renormalisation of four-fermi terms, involving Weyl fermions, in Appendix \ref{beta}.

\section{Chiral Gauge Theories}
\label{p0theories}

In this section, we study a pair of two dimensional chiral gauge theories. Each is built around an $SU(N)$
gauge group, with a Weyl fermion in either the symmetric or anti-symmetric representation,
together with a collection of further Weyl fermions in the anti-fundamental. 

We start in Section \ref{ssAS} by introducing the first of the chiral gauge theories which contains an anti-symmetric Weyl fermion. We present its symmetry and anomaly structure, apply the coset proposal and compare this to a different low energy phase composed of free fermions. In Section \ref{symmetric}, we present all of the above for a second very similar theory which contains a Weyl fermion in the symmetric representation of the gauge group.

\subsection{$SU(N)$ with an Anti-Symmetric}
\label{ssAS}
Consider an $SU(N)$ gauge theory with a single right-moving Weyl fermion $\tchi$ in the $\anti$ and $q$ left-moving fermions $\psi$ in the $\overline{\Box}$. Gauge anomaly cancellation requires
\be
q= N-2
\nn\ee
which follows from the Dynkin indices $\mu(\,\anti\,)= N-2 $, $\mu(\Box) =1$ for $SU(N)$ together with $\mu(R)= \mu (\bar{R})$. This is the 2d analogue of certain 4d chiral theories, with an anti-symmetric fermion, which have been extensively studied in the literature \cite{appel,konishi,mury1,mury2,us,avner,me,zepp}. This chiral theory has a continuous global symmetry group
\be
G=SU(N-2)\times U(1)_R \times U(1)_L
\nn\ee
Discrete quotients do not play a role in our story so we omit them throughout the main text but discuss them in Appendix \ref{discrete}. Under the combination $SU(N)\times G$, the full collection of fermions transforms as
\be
\begin{array}{|c||c||c|c|c|c|c|c|}
\hline & SU(N) & SU(N-2) & U(1)_{R} & U(1)_L \\ \hline
\rule{0pt}{2.5ex} 
\tchi & \anti & \mathbf{1} & 1 & 0 \\
\rule{0pt}{2.5ex} 
\psi & \overline{\Box} & \Box & 0 & 1 \\
\hline
\end{array}\label{antisym0}\ee
We will refer to this as the \emph{anti-symmetric theory}. Here and throughout this paper we denote right-handed fermions with a tilde and left-handed ones without.

Anomalies constrain the IR physics. It is straightforward to compute these.
\be
\begin{aligned}
&{\cal A}[SU(N-2)^2]=N\\
&{\cal A}[U(1)_R^2]=-\dfrac{N}{2}(N-1)\\
&{\cal A}[U(1)_L^2]= N(N-2)\\
&{\cal A}[U(1)_R\cdot U(1)_L]=0\\
\end{aligned}
\label{asanom}
\ee
The theory also has a gravitational anomaly
\be
{\cal A}[\text{Grav}]=c_L-c_R= \dfrac{N}{2}(N-3)
\label{asanomgrav}
\ee
and, as discussed in Appendix \ref{discrete}, there are no further discrete anomalies. These must all be reproduced in the IR. So what is the low energy physics of this theory? There are two candidates which have been proposed in the literature.

The first option, proposed in \cite{Delmastro,Delmastro2}, is that this theory flows to a certain coset CFT. The argument for this phase is natural and intuitive. Consider a 2d gauge theory with a compact, connected gauge group $G$ and left/right-handed Weyl fermions, $\psi_l$ and $\tilde{\psi}_r$, in representations $R_l$ and $R_r$ of $G$ respectively. The action is
\be
S=\int d^2x\,\left( -\frac{1}{2g^2} \operatorname{tr}\left(F_{\mu \nu} F^{\mu \nu}\right)+i \psi_{\ell}^{\dagger} D_{-} \psi_{\ell}+i {\tilde{\psi}}_{r}^{\dagger} D_{+} \tilde{\psi}_{r}\right)
\nn\ee
with the usual field strength and covariant derivatives in light-cone coordinates. Now observe that the Yang-Mills coupling $g$ has positive mass dimension such that the gluon kinetic term is classically irrelevant. If we assume this continues to hold in the full quantum theory, $g\rightarrow \infty$ as $E\rightarrow 0$, the kinetic term vanishes under RG flow. The intuition here is that 2d gluons do not propagate. This leaves the fermionic terms $\mathcal{L}_{\mathrm{eff}}\sim \psi^\dag D \psi$. The gauge fields act as Lagrange multipliers and set the gauge currents to zero. Then bosonising leads to the coset CFT/TQFT \cite{WittenNA,Kutasov,Goddard1,Komargodski}
\be
\mathcal{T}_{\text{IR}}=\frac{U\left(\operatorname{dim} R_{\ell}\right)_{1}}{G_{\mu\left(R_{\ell}\right)}} \times \frac{U\left(\operatorname{dim} R_{r}\right)_{1}}{G_{\mu\left(R_{r}\right)}}
\nn\ee
We call this the \emph{coset proposal}. The vanishing of the kinetic term is of course a first guess\footnote{Furthermore, as mentioned in \cite{Komargodski}, the $g \rightarrow \infty$ limit is not strictly meaningful as $g$ is the only scale in the theory. Nevertheless, the coset proposal is a natural candidate for the low-energy theory.}. It is possible  that classically irrelevant terms pick up large quantum corrections, become ``dangerously irrelevant" and alter the low-energy physics. Nonetheless, the coset proposal is a sensible IR phase.

The coset model factorises into independent chiral and anti-chiral halves. This is a consequence of assuming that the Yang-Mills kinetic term vanishes in the IR limit. Once we drop this term, the remaining gauge field components $A_\pm$ act as independent Lagrange multipliers for left/right-handed fermions and do not mediate any interactions between them. This means any symmetry of the UV gauge theory which is purely holomorphic remains holomorphic in the IR, and similarly for the anti-holomorphic currents. There is no mixing of left and right-moving symmetries.

Applying this proposal to the the anti-symmetric theory gives our first candidate IR CFT
\be
\mathcal{T}=\dfrac{U (N(N-2))_{1}}{SU(N)_{N-2}} \otimes  \left[ \dfrac{U(N(N-1)/2)_1}{SU(N)_{N-2}} \right]\cong U(N-2)_N \otimes U(1)_{N(N-1)/2}
\nn\ee
The equivalence follows from the branching rules of the cosets. These are known by level-rank duality $\mathfrak{g}(N)_{N_f} \leftrightarrow \mathfrak{g}(N_f)_{N} $. Writing out the $U(1)_L$ factor explicitly gives
\be
\mathcal{T}= \left(SU(N-2)_N \otimes U(1)_{N(N-2)}\right)\otimes U(1)_{N(N-1)/2}
\label{eq:BYcoset0}\ee
By construction, the CFT matches the symmetry structure of the UV theory and reproduces the anomalies \eqref{asanom} and \eqref{asanomgrav}. This can simply be seen from the levels of the WZW models. 

There is a second, rather natural candidate for the IR physics proposed by Tong \cite{DTsm}. The theory may confine, reducing to the following gauge singlet fermions in the IR:
\be
\lambda = \psi \tchi \psi,\;\;\;\;  \tilde{\mathcal{B}} \sim \psi^{N-2}\tchi^{N-1}
\nn\ee
Fermi statistics fix $\lambda$ to have symmetric flavour indices. The baryonic operator $\tilb$ can be constructed as
\be
\tilb = \epsilon_{a_1...a_{N-2}} \psi_{i_1}^{a_1}...  \psi_{i_{N-2}}^{a_{N-2}}\epsilon_{j_1...j_N} \tchi^{i_1j_1}... \tchi^{i_{N-2}j_{N-2}}\tchi^{j_{N-1}j_N}
\nn\ee
where $i,j=1,...,N$ indices are associated with the $SU(N)$ gauge group and $a=1,...,N-2$ indices with the $SU(N-2)$ global symmetry. $\lambda$ is left-moving and $\tilb$ is right-moving due to the spins of the constituent fermions. Under the global symmetry $G$, the quantum numbers of the fermions are
\be
\begin{array}{|c||c|c|c|c|c|c|}
\hline  & SU(N-2) & U(1)_{R} & U(1)_L \\ \hline 
\lambda & \sym & 1 & 2  \\
\tilde{\mathcal{B}} & \mathbf{1} & N-1 & N-2 \\
\hline
\end{array}\label{sconfanti}\ee
It is again straightforward to check that these reproduce the anomalies \eqref{asanom} and \eqref{asanomgrav}.

How do these two proposals compare? The central charges of the two low-energy theories are identical
\be
c_\text{coset}=c_\text{fermion} = \dfrac{N}{2}(N-3)+2\equiv c_\text{IR}
\nn\ee
with $c_\text{IR} \leq c_\text{UV}$ for all $N\geq2$ (only the $N=2$ case saturates the bound as the UV theory reduces to a single free right-handed fermion with no flow). It also is known that the $SU(N-2)_N$ \linebreak WZW model has a free fermion representation in terms of complex fermions in the $\sym$ of $SU(N-2)$ \cite{Goddard}. This can simply be seen by constructing the current $J^A(z)= \operatorname{tr}(\lambda^\dag T^A_{\scalebox{0.5}{\sym}}\lambda )(z)$, where $T^A_{\scalebox{0.5}{\sym}}$ is the symmetric representation generator, which reproduces the current algebra. 

So the two theories share the same central charge and the $SU(N-2)$ chiral algebras match. Are the two theories identical?

Almost, but not quite. The difference is subtle and appears in the factorisation of the $U(1)_L\times U(1)_R$ symmetries. In the coset proposal, the (anti)holomorphic current of the UV theory descends to a purely (anti)holomorphic current in the IR. The left-right decomposition of $U(1)$ charges remain preserved under RG flow.

On the contrary, the fermion phase does not retain the left-right decomposition. There is charge mixing, where the IR left-mover $\lambda$ carries charges under both $U(1)_R\times U(1)_L$ symmetries, and similarly for the IR right-mover $\tilb$.

As a consequence, the physics of the two IR proposals differ. This is intuitively clear from the following argument.\footnote{I thank Avner Karasik for pointing this out to me} Consider creating a wave packet charged only under $U(1)_L$, and turn on a background gauge field for $U(1)_L$. In the coset proposal, the wave packet will stay purely left-moving, whilst in the fermion proposal, the wave packet will split into right and left-moving parts.

\subsubsection*{A Conformal Manifold and RG Flows}
So how are the coset and massless fermion models related? They both have the same symmetry structure, which includes $U(1)_R\times U(1)_L$. As a consequence, they may both be deformed by the exactly marginal current-current interaction
\be
\mathcal{L}^\text{IR} = \tilde{a} J\bJ
\label{jjbar}\ee
On the free fermion side, this corresponds to the four-fermi term $\mathcal{L}_4^\text{IR}= \tilde{a} \operatorname{tr}(\lambda^\dag \lambda) (\tilb^\dag \tilb)$. This gives rise to a one-dimensional conformal manifold $\mathcal{M}_\text{IR}$ of low-energy theories parameterised by $\tilde{a}$. We propose that the coset and free fermion proposals correspond to different points on $\mathcal{M}_\text{IR}$. They can be continuously connected to one another by moving along this moduli space of conformal field theories.

As the conformal manifold $\mathcal{M}_\text{IR}$ is parameterised by current-current interactions, its origin can be traced back to the exactly marginal four-fermion deformation
\be
\mathcal{L}_4^\text{UV} = a \operatorname{tr}(\tchi^\dag \tchi) \operatorname{tr}(\psi^\dag \psi) 
\label{UV4f}\ee
of the UV  chiral gauge theory \eqref{antisym0}. We propose that as we vary $a$, we land on different point on $\mathcal{M}_\text{IR}$.

So how does the four-fermi deformation \eqref{UV4f} of the UV theory affect the RG flow? Let us start with the $a=0$ case. The only interactions are due to the gauge fields. If $g\rightarrow \infty$ in the deep IR, this flows to the coset \eqref{eq:BYcoset0}. We assume that this is the correct low-energy limit.

Now we turn on the four-fermi deformation \eqref{UV4f}. For small $a\ll 1 $, this maps to a $J\bar{J}$ perturbation of the IR coset. Such deformations of WZW models are well understood \cite{Forste,Forste2,Delmastro2}. In terms of the IR conformal manifold $\mathcal{M}_\text{IR}$, this corresponds to a small step away from the coset point to a point on $\mathcal{M}_\text{IR}$ near it.

As we now increase $a$, we move smoothly along $\mathcal{M}_\text{IR}$  and flow to different points further away from the coset. We propose that as $a\rightarrow \infty$, the UV theory flows to the free, confined fermions \eqref{sconfanti}. A finite but large $a$ then moves us away from the free fermion point by turning on the marginal four-fermion deformation $\delta \mathcal{L}_4^\text{IR} \sim
 \frac{1}{a} \operatorname{tr}(\lambda^\dag \lambda) (\tilb^\dag \tilb)$. 
 
\begin{figure}[h]
	\centering
	\includegraphics[width=0.55 \linewidth]{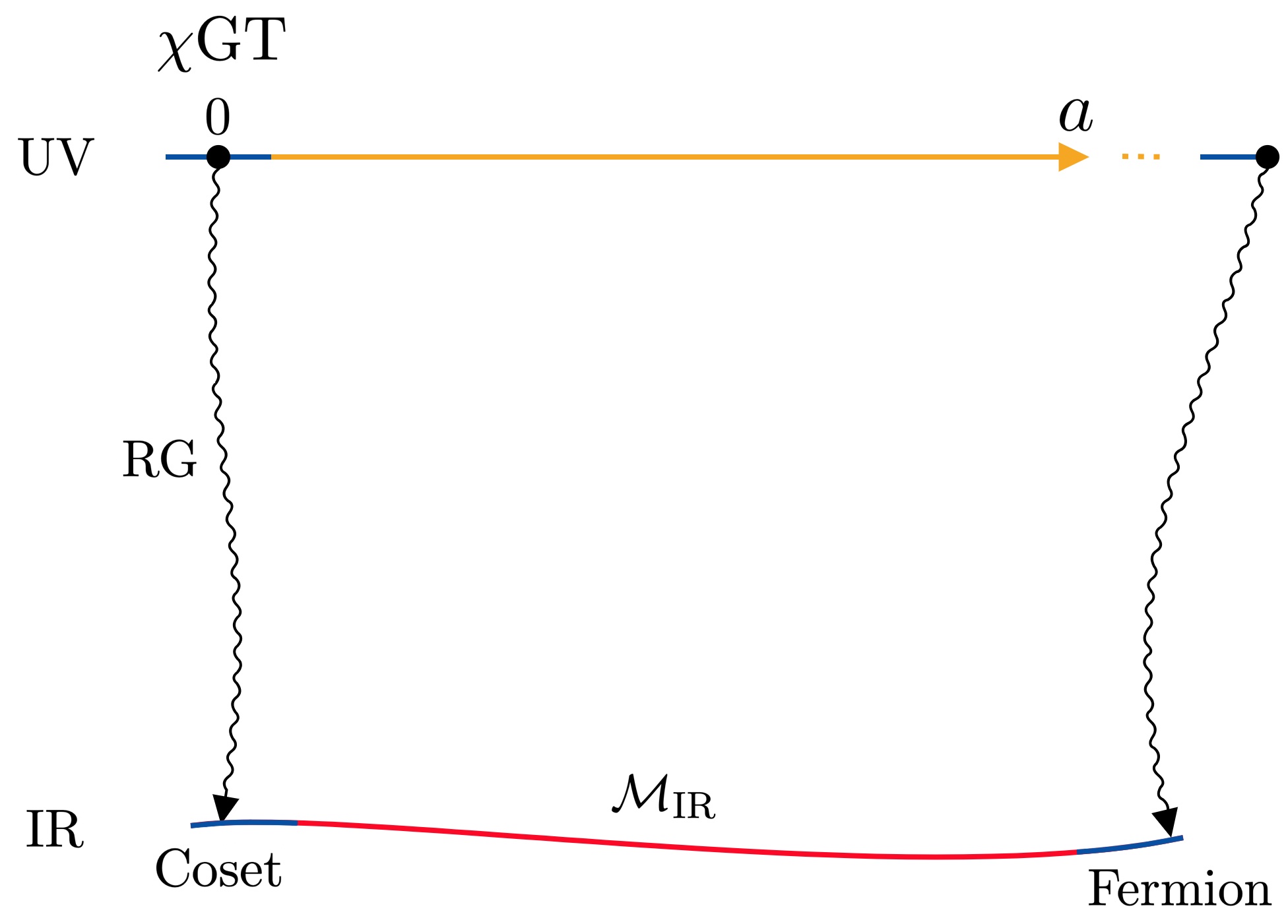}
	\caption{Proposed RG flows of the UV chiral gauge theory ($\chi$GT) as it is deformed by the Abelian current-current interaction $\mathcal{L}_4^\text{UV} = a J \bar{J}$. At $a=0$, the theory flows to the coset and as $a\rightarrow\infty$ it flows to free fermions. Both of these theories are proposed to live on the same conformal manifold $\mathcal{M}_\text{IR}$. The UV (blue) regions in the vicinity of these fine tuned points correspond to small $J\bJ$ deformations of the IR theories.}
	\label{flow}
\end{figure}

 This is a natural proposal as these are the only possible deformations of the theories consistent with symmetries. All anomalies match and both IR theories have the same central charge.  The two low-energy theories exhibit a strong-weak duality, as a large $\sim \tilde{a}J\bJ$ operator deforming the coset is equivalent to a small $\sim \frac{1}{\tilde{a}}J\bJ$ term deforming the free fermions. We present the RG flow schematically in Figure \ref{flow}.

At this point, some subtleties need to be addressed. First, the deformation \eqref{jjbar} is written with a finite coupling constant and the currents $J$ and $\bJ$ of the original unformed theory. It is instead also possible to consider the deformation
\be
\dfrac{\partial\mathcal{L}(\mu)}{\partial\mu}= J(\mu)\bJ(\mu)
\label{jjbar2}\ee
where this time the currents are of the deformed theory with $J(0)=J$ etc., so one might worry that this is the correct deformation to study. In fact, $T\bar{T}$ deformations with analogous definitions are known to differ significantly \cite{zomo}. For our simpler $J\bJ$ case however, the two deformations \eqref{jjbar} and $\eqref{jjbar2}$ coincide. This is because the only effect of the $J\bJ$ deformation is to rescale the currents such that $J(\mu)\bJ(\mu)=f(\mu)J\bJ$ for some function $f$ (see the appendix of \cite{Bzowski} for details) and the effects of the two deformations are identical.

Second, it is also possible to consider a single-trace deformation
\be
\tilde{\mathcal{L}}_4^\text{UV} = b \operatorname{tr}(\psi\tchi \tchi^\dag \psi^\dag) 
\nn\ee
of the UV theory which corresponds to a contraction of gauge-currents and will get renormalised. For small $b\ll 1$, the effect of this deformation on the low energy physics is identical to the previously considered \eqref{UV4f}. This simply follows as the gauge dynamics kick in which set the traceless part of the gauge currents to vanish, leaving $\sim \frac{b}{N} \operatorname{tr}(\tchi^\dag \tchi) \operatorname{tr}(\psi^\dag \psi) $. Further details of this are discussed below in \eqref{gaugecurrentfinal}. It is not possible to be concrete about large $b$ but this does not affect our discussion.

Finally, we should understand why the free fermion point corresponds to a point asymptotically far out on the moduli space $\mathcal{M}_\text{IR}$. We will present further evidence for this below but there is also a simple argument demonstrating that this should be expected.  We should add interaction terms to the Hamiltonian that are positive definite. This puts a lower bound on the $\tilde{a}$ coupling of the free fermions, as was originally discussed by Coleman for the Thirring model \cite{coleman2}. If the free fermions were to sit at some finite value of $\tilde{a}$, one direction on $\mathcal{M}_\text{IR}$ would point towards the coset but the other direction would take us further away on the moduli space. However, the sign of the four-fermion coupling would be different for the two directions, meaning that the fermion theory would become unbounded from below in one direction. This makes it very reasonable that the free fermion point sits at the end of the moduli space with a single consistent direction away from it.

\subsubsection*{Current Mixing From $J\bar{J}$ Deformations}
\label{chargemix}
The mixing of $U(1)$ charges plays an important role for this proposal to be consistent. If the two theories are to correspond to different points on $\mathcal{M}_\text{IR}$, the deformation $J\bJ$ must rotate left/right currents and cause mixing. Only then can we start from the coset theory, with diagonal charges, and deform to the fermion phase with charge mixing and vice versa.

To this end, we present a simple CFT argument demonstrating that a perturbative $J\bJ$ deformation leads to a mixing of $U(1)$ currents. Similar arguments were previously considered in the appendix of \cite{Bzowski}. Consider a CFT with a left-moving $U(1)_L$ symmetry $J$ and a right moving $U(1)_R$ symmetry $\bJ$. These necessarily obey the factorised conservation laws $\bar{\partial}J=\partial \bJ =0$. It follows that any anti-holomorphic operator $\bO(\bar{w})$ can only be charged under $U(1)_R$ and must be neutral under $U(1)_L$ (we focus on anti-holomorphic operators for the rest of this section. Identical statements hold for the holomorphic sector simply by swapping $U(1)_L$ and $U(1)_R$.) In terms of OPEs, this translate to
\be
\bJ(\bar{z})\bO(\bar{w})\sim \dfrac{\bar{q}\bO(\bar{w})}{\bar{z}-\bar{w}}, \;\;\;\; J(z)\bO(\bar{w}) \sim 0
\label{ope}\ee
for $\bO$ with charge $\bar{q}$ under $U(1)_R$.

We now deform the action by the exactly marginal current-current operator 
\be
S_0\rightarrow S= S_0 - \epsilon \int d^2z \,J(z)\bJ(\bar{z})
\nn\ee
and ask if this leads to charge mixing. The relevant correlation function to consider is
\be
\langle J(z) \bO(\bar{y}) \rangle_{\epsilon \neq 0} =\langle J(z) \bO(\bar{y}) e^{\epsilon\int d^2 w \, J \bJ} \rangle_{ 0}
\nn\ee
where $\langle ...\rangle_{ 0}$ are the correlators of the original, undeformed theory. If this is nonvanishing, $\bO$ has picked up a charge under $U(1)_L$. Expanding in small $\epsilon\ll1$, we get
\be
\begin{aligned}
\langle J(z) \bO(\bar{y}) \rangle_{\epsilon \neq 0} &= \epsilon \langle J(z) \bO(\bar{y}) \int d^2 w \, J(w) \bJ(\bar{w}) \rangle_{ 0} + \mathcal{O}(\epsilon^2)\\ 
&= -  2 \pi k \bar{q} \epsilon \dfrac{\bO(\bar{y})}{z-y} + \mathcal{O}(\epsilon^2)
\end{aligned}
\label{opec}\ee
which follows from \eqref{ope} and
\be
J(z)J(w)\sim \dfrac{k}{(z-w)^2}\;\;\;\;\;\; \text{ and} \;\;\;\;\;\; \dfrac{\partial}{\partial \bar{z}}\left(\dfrac{1}{z-w}\right)=2\pi \delta^2(z-w)
\nn\ee
We see that the deformation leads $\bO$ to pick up a charge under $U(1)_L$. 

The charge in \eqref{opec} is not quantised. This seems problematic but is actually to be expected. The non-integer charges only appear with respect to the original holomorphic current $J$ and it instead possible to define a non-chiral current whose charges remain conserved as the deformation is turned on \cite{Bzowski}. This can be viewed as a choice of active versus passive transformation and is consistent with our proposal. As the deformation is turned on, anti-holomorphic operators can pick up charge under $U(1)_L$ with the original current $J$.

\subsubsection*{Continuously Connecting the Proposals}
\label{relatingproposals}

We now exhibit a continuous path, preserving symmetry, from the UV gauge theory with no four-fermi terms, to the IR free fermion phase. This is the 2d analogue of the well-known 4d ``complementarity" arguments considered in \cite{lcf,us,seibergtalk} and continuously connects the coset and free fermion proposals.

We start by writing the gauge and flavour indices of the fermions explicitly; these are $\tchi_{ij}$ and $\psi^i_a$, where $i,j = 1, ..., N$ are the gauge indices and $a = 1, ..., N-2$ is the flavour index under the $SU(N-2)$ global symmetry.

We then extend the UV theory by introducing $N-2$ anti-fundamental scalars $\phi^{ia}$ coupled to the fermions by the Yukawa term
\be
\mathcal{L}_Y=y \phi \psi \tchi +\text{h.c.}
\nn\ee
where the coupling $y$ has a dimension of mass. The addition of the scalar does not change the symmetry structure, including the discrete quotients discussed in Appendix \ref{discrete}. However, the $U(1)$ currents do get modified due to the scalars and are no longer holomorphic. This will be crucial to reach the free fermion phase. Together with the fermions, the quantum numbers of the fields read
\be
\begin{array}{|c||c||c|c|c|c|c|c|}
\hline & SU(N) & SU(N-2) & U(1)_{R} & U(1)_L \\ \hline
\rule{0pt}{2.5ex} 
\tchi & \anti & \mathbf{1} & 1 & 0 \\
\rule{0pt}{2.5ex} 
\psi & \overline{\Box} & \Box & 0 & 1 \\
\hline\
\rule{0pt}{2.5ex} 
\phi &\overline{\Box} &\overline{\Box} & -1 & -1\\
\hline
\end{array}\label{uvtheory}\ee

The gauge and Yukawa couplings, $g$ and $y$, have mass dimension and as the mass $m_\phi$ of the scalar varies, the
theory has two different phases. In the $m^2_\phi \gg g^2 $ and $m^2_\phi \gg y^2 $ phase, the scalars simply decouple and we are left with the original gauge theory. We assume this then flows to the coset \eqref{eq:BYcoset0}.

This decoupling is strictly true in the $m_\phi^2/g^2\rightarrow \infty$ and $m_\phi^2/y^2\rightarrow \infty$ limits. For $m_\phi$ much larger than the other couplings but finite, integrating out the massive scalars will lead to effective interactions of the low energy theory. To track such leading-order effective interactions, we probe the theory at energy scale $E$ with $m_\phi^2 \gg y^2 \gg E^2\gg g^2$. Integrating out the scalars results in the gauge current-current deformation
\be
\mathcal{L}^E_\text{eff}= \dfrac{y^2}{2m_\phi^2}\operatorname{tr}(J \bJ)
\label{def1}\ee
of the original anti-symmetric theory \eqref{antisym0} without scalars. Here the gauge current in holomorphic coordinates reads
\be
J^i_j= -\psi^{\dag a}_{j} \psi^{i}_a ,\;\;\; \bar{J}^i_j = 2\tchi^{\dag i k}\tchi_{jk}
\label{gaugecurrent}\ee
which follows from the relation between the anti-symmetric and fundamental generators $(T^A_{\scalebox{0.5}{\anti}})^{ij}_{\,\,\; kl}=2 (T^A_{\scalebox{0.5}{\ydiagram{1}}})^{[i}_{\,\; [k} \delta^{j]}_{\,\; l]}$. Now, as we flow to lower energies $g^2\gg \tilde{E}^2$ the gauge dynamics kick in. Assuming that it is irrelevant, we drop the gauge kinetic term and the remaining gauge-fermion interaction $\operatorname{tr} (A^\mu J_\mu)$ sets the traceless parts of \eqref{gaugecurrent} to zero. This reduces \eqref{def1} to the deformation
\be
\mathcal{L}^{\tilde{E}}_\text{eff}= \dfrac{y^2}{2N m_\phi^2}\operatorname{tr}(J) \operatorname{tr}(\bJ)
\label{gaugecurrentfinal}\ee
which is simply the exactly marginal $U(1)$ current-current deformation introduced in \eqref{UV4f}. In the deep IR, we end up with the coset \eqref{eq:BYcoset0} deformed by a small $U(1)$ current-current deformation $\sim J\bJ$.

Another regime, $m^2_\phi \ll - g^2$, is the Higgs phase. We construct a potential to form a diagonal vacuum expectation value
\be
\langle \phi^{ia} \rangle = v\delta^{ia}=\left(\begin{array}{ccc}
v & \ldots & 0 \\
& \ddots & \\
0 & \ldots & v \\
\hdashline 0 & \ldots & 0\\
0 & \ldots & 0
\end{array}\right)
\nn\ee
In terms of the scalar Lagrangian, the vacuum expectation value is set by $v^2=|m_\phi|^2/\lambda$, where $\lambda$ is the $|\phi|^4$ coefficient. The effect of this vacuum is to Higgs the gauge group $SU(N)\rightarrow SU(2)$. The global symmetries are left intact, albeit after locking with the gauge symmetry. To see this, we write the most general $SU(N)\times SU(N-2)\times U(1)_L\times U(1)_R$ transformation acting on $\phi$ as
\be
\phi \rightarrow e^{-2\pi i (\alpha +\tilde{\alpha})} U^* \phi V^\dag
\nn\ee
with $U \in SU(N)$ and $V\in SU(N-2)$. The choice
\be
U=\begin{pmatrix}
V^* e^{-2\pi i (\alpha +\tilde{\alpha})} & 0\\
0 & U' e^{2\pi i (\alpha +\tilde{\alpha})(N-2)/2}
\end{pmatrix}
\nn\ee
with $U'\in SU(2)$ leaves the vacuum invariant. The global symmetries remain (as they must in 2d) but the gauge group has been Higgsed down to $SU(2)$. To analyse the $U(1)$ charges more closely, we take the linear combinations $Q_V=(Q_L-Q_R)/2$ and $Q_A=(Q_L+Q_R)/2$ such that the fields have the charges
\be
\begin{array}{|c||c||c|c|c|c|c|c|}
\hline & SU(N) & SU(N-2) & U(1)_{V} & U(1)_A \\ \hline
\rule{0pt}{2.5ex} 
\tchi & \anti & \mathbf{1} & -\frac{1}{2} & \frac{1}{2} \\
\rule{0pt}{2.5ex} 
\psi & \overline{\Box} & \Box & \frac{1}{2} & \frac{1}{2} \\
\hline
\rule{0pt}{2.5ex} 
\phi & \overline{\Box} & \overline{\Box} & 0 & -1 \\
\hline
\end{array}\nn\ee
We further define an unbroken axial symmetry $U(1)'_A$ by mixing  $U(1)_A$ with a suitable generator of the $SU(N)$ gauge symmetry
\be
Q'_A=Q_A+\dfrac{1}{2}\operatorname{diag}(\underbrace{-2,...,-2}_{N-2},N-2,N-2)
\nn\ee
such that $Q'_A[\langle \phi \rangle]=0$.

In this Higgs phase, all scalars are gapped. The Yukawa interaction results in a mass term of the form $v y \sum_{i=1}^N\sum_{a=1}^{N-2}\psi_{ia}\tchi_{ia}$. This gives a mass to most of the fermions. The only ones that survive in the IR transform under $SU(N)\times SU(N-2)\times U(1)_V\times U(1)'_A$ as
\be
\begin{aligned}
\text{left-moving } \lambda \text{ in } (\mathbf{1},\sym)_{(1/2,3/2)}: \text{ this comes from } \psi \\
\text{right-moving } \tilb \text{ in } (\mathbf{1},\mathbf{1})_{(-1/2,N-3/2)}: \text{ this comes from } \tchi
\end{aligned}\label{IRfermions} \ee
with the remaining $SU(2)$ gauge group decoupling. Crucially, $\tilb$ is in the anti-symmetric representation of $SU(2)$ but this is a singlet. Translating the $U(1)$ charges back to the original $U(1)_L\times U(1)_R$ basis, we recover the massless, free fermions in \eqref{sconfanti}:
\be
\begin{array}{|c||c|c|c|c|c|c|}
\hline  & SU(N-2) & U(1)_{R} & U(1)_L \\ \hline 
\lambda & \sym & 1 & 2  \\
\tilde{\mathcal{B}} & \mathbf{1} & N-1 & N-2 \\
\hline
\end{array}\nn\ee
We find that the massless spectrum consists of massless fermions. These fermions are free in the strict $m_\phi^2/g^2\rightarrow - \infty$ limit and any effective interaction generated must be suppressed by $g^2/|m_\phi|^2$.

Such effective interactions are generated by integrating out the heavy gauge fields. The scalar vacuum expectation value gaps most of the gauge fields by the term
\be
-v^2\sum_{i=1}^N\sum_{a=1}^{N-2}(A^\mu)^i_a (A_\mu)^a_i
\nn\ee
The only remaining massless gauge fields $(A^\mu)^I_J$ with indices $I,J\in\{N-1,N\}$ are those of the unbroken $SU(2)$. The gauge fields are also coupled to the fermions by the usual $\operatorname{tr} (A^\mu J_\mu)$ interaction with the gauge currents \eqref{gaugecurrent}. The fact that the generators of the $\mathfrak{su}(N)$ Lie algebra are traceless leads to  $\operatorname{tr} (A^\mu J_\mu) = \operatorname{tr} (A^\mu \mathcal{J}_\mu)$, where $\mathcal{J}_\mu = J_\mu - I_N \operatorname{tr}(J_\mu)/N$ is the traceless part of the gauge current. From here, we can integrate out the heavy gauge fields which results in current-current interactions
\be
\mathcal{L}_\text{eff}=-\dfrac{1}{2v^2}\left( \mathcal{J}^a_b \overline{\mathcal{J}}^b_a + \mathcal{J}^a_I \overline{\mathcal{J}}^I_a\right) +\text{h.c.}
\nn\ee
with $a,b=1,...,N-2$ and $I=N-1,N$. Among these terms is a current-current interaction for the massless fermions,
\be
\mathcal{L}_\text{eff} \supset \dfrac{1}{N^2 v^2} \left(\psi^\dag_{[ab]}\psi^{[ab]}\right)\left( \tchi^{\dag \,IJ} \tchi_{IJ}\right) \equiv \dfrac{1}{2 N^2 v^2} \operatorname{tr}( \lambda^\dag \lambda ) (\tilb^\dag \tilb)
\label{marginal}\ee
where $\lambda^{ab}=\psi^{[ab]}$ and $\tilb =\epsilon^{IJ}\tchi_{IJ}$ are the massless IR fermions written above in \eqref{IRfermions}.

We find that in the Higgs phase the massless fields do not sit at the free fermion point, except when $m^2_\phi/g^2\rightarrow - \infty$. Instead, the gauge interactions induce the exactly marginal coupling \eqref{marginal} on the low energy theory
and we have an interacting CFT in IR.

This fits well with the dynamics and IR conformal manifold proposed above. In the strict $m_\phi^2/g^2\rightarrow - \infty$ limit we hit free fermions. As we then vary the dimensionless ratio $m^2_\phi/g^2$, this turns on the marginal deformation \eqref{marginal} and we trace a path in the moduli space of conformal field theories $\mathcal{M}_\text{IR}$. As we reach the opposing $m_\phi^2\gg g^2$ limit, we have the coset theory \eqref{eq:BYcoset0} deformed by the $U(1)$ current-current interaction $J\bJ$. Finally in the strict $m_\phi^2/g^2\rightarrow \infty$ limit, we hit the undeformed coset model. By analysing the extended UV theory \eqref{uvtheory}, we have thus continuously connected the free fermion and coset phases.

We finally note that in \cite{DTsm} Tong originally conjectured free fermions as the end point of the pure gauge theory with no Higgs fields. However, the Higgsing argument of this section provides good evidence that this is not the case. The free fermion theory arises in the deep Higgs phase. Then as we go back to the un-Higgsed phase, we deform the free fermions by a four-fermion term and end up with the coset as the IR limit of the un-Higgsed theory.

\subsection{$SU(N)$ with a Symmetric}
\label{symmetric}

We now move on to consider a similar $SU(N)$ chiral gauge theory with a  right-moving Weyl fermion in the $\sym$ and $N+2$ left-moving fermions in the $\overline{\Box}$. The Dynkin index $\mu(\,\sym\,)= N+2 $ ensures that the theory is non-anomolous. Ignoring any discrete subtleties, the global symmetry group is 
\be
G=SU(N+2)\times U(1)_R \times U(1)_L
\nn\ee
under which the fermions transform as
\be
\begin{array}{|c||c||c|c|c|c|c|c|}
\hline & SU(N) & SU(N+2) & U(1)_{R} & U(1)_L \\ \hline 
\tilde{\chi} & \sym & \mathbf{1} & 1 & 0 \\
\psi & \overline{\Box} & \Box & 0 & 1 \\
\hline
\end{array}\label{symUV}\ee
We refer to this as the \emph{symmetric theory}. It is again possible to turn on the exactly marginal deformation
\be
\mathcal{L}_4^\text{UV} = a \operatorname{tr}(\tchi^\dag \tchi) \operatorname{tr}(\psi^\dag \psi) 
\nn\ee
and our goal is to understand the low energy physics as a function of $a$. There are again two candidates.

Setting $a=0$, the coset proposal \eqref{QCDcoset1}  gives
\be
\mathcal{T}=\dfrac{U (N(N+2))_{1}}{SU(N)_{N+2}} \otimes  \left[ \dfrac{U(N(N+1)/2)_1}{SU(N)_{N+2}} \right]\cong U(N+2)_N \otimes U(1)_{N(N+1)/2}
\nn\ee
where the $U(1)_L$ level is explicitly $U(N+2)_N \equiv SU(N+2)_N \times U(1)_{N(N+2)}$. Again, by construction, we have decoupled chiral and anti-chiral CFTs with no charge mixing.

Option two is the confined, massless gauge-singlet fermions proposed by Tong \cite{DTsm}. These are
\be
\lambda = \psi \tchi \psi,\;\;\;\;  \tilde{\mathcal{B}} \sim \psi^{N+2}\tchi^{N+1}
\nn\ee
This time fermi statistics fix $\lambda$ to have anti-symmetric flavour indices and the baryonic state $\tilb$ can be constructed as
\be
\tilb = \epsilon_{a_1...a_{N+2}} \psi_{i_1}^{a_1}...  \psi_{i_{N+2}}^{a_{N+2}}\epsilon_{j_1...j_N} \tchi^{i_1j_1}... \tchi^{i_{N}j_{N}}\partial_{-}\tchi^{i_{N+1}i_{N+2}}
\nn\ee
where the $\partial_{-}$ is necessary to ensure that $\tilb$ is right-moving even though its constituents include one more left-mover than right-mover. It is straightforward to check that the anomalies match.

Much of the story we presented for the anti-symmetric theory repeats itself for the symmetric case. The central charges of the IR proposal are again identical $c_\text{IR}=N(N+3)/2+2<c_\text{UV}$ for all $N\geq 2$. The symmetries and anomalies match. Again, the only obstruction to bosonising/fermionising the two low energy theories to one another are the $U(1)$ levels. The fermions have charge mixing under RG flow, whereas the coset remains factorised.

We propose that the RG flows of the symmetric theory take the form we have presented for the anti-symmetric theory (see Figure \ref{flow}). The $a=0$ case is assumed to flow to the coset. Then, as $a\rightarrow \infty$, the symmetric theory flows to free fermions. The two low energy theories sit on different ends of a one-dimensional conformal manifold $\mathcal{M}_\text{IR}$.

\subsubsection*{An Extended UV Theory}
For the symmetric theory, the Higgs and confined fermion phases differ and we can not repeat the complementarity argument. Instead, in this section we construct a path in field space that continuously connects the fermions and the coset. The result is not as strong as the previous complementarity argument but shows that the free fermions and the coset are in the same deformation class \cite{seibergtalk}. The construction we present is the 2d analogue of the 4d construction in Section 3.2 of \cite{us}.

To achieve this, we start with a different UV theory. We show that as we dial
some parameters, we can interpolate between the symmetric chiral gauge theory, and the massless free fermions.

The parent UV theory we consider is a $G=SU(N+2)\times SU(N)$ gauge theory with the matter content
\be
\begin{array}{|c||c|c||c|c|c|c|c|c|}
\hline & SU(N+2) & SU(N) & SU(N+2) & U(1)_1 & U(1)_2 \\ \hline
\rule{0pt}{2.5ex} 
\lambda & \anti & \mathbf{1} & \mathbf{1} & 1 & 0 \\
\rule{0pt}{2.5ex} \tilde{\psi} & \overline{\Box} & \Box & \mathbf{1} & 0 & 1 \\
\eta & \mathbf{1} & \overline{\Box} & \Box & 0 & -1 \\
\tilb   & \mathbf{1} & \mathbf{1} & \mathbf{1} & N+1 & N \\ \hline
\rule{0pt}{2.5ex} 
\phi & \overline{\Box} &  \overline{\Box}  & \mathbf{1} & -1 & -1 \\
\theta & \Box & \mathbf{1} &\overline{\Box} & 0 & 0 \\ \hline
\end{array}\nn\ee
The first four rows are fermions and the bottom two rows are scalars. The symmetry structure follows from the interaction terms
\be
\mathcal{L} \sim \tilde{\psi}^{N+2} \eta^{N+2} + \phi^{N} \lambda^\dag \tilb +\phi \tilde{\psi} \lambda +\theta \tilde{\psi}\eta + \text{h.c.} 
\label{int}\ee
The indices of the first two terms are explicitly
\be
\begin{aligned}
\tilde{\psi}^{N+2} \eta^{N+2}=& \epsilon^{i_1...i_{N+2}}\epsilon_{a_1...a_{N+2}}({\tilde{\psi}}_{i_1}^{I_1} \eta_{I_1}^{a_1})...(\tilde{\psi}_{i_{N+2}}^{I_{N+2}}\eta_{I_{N+2}}^{a_{N+2}})\\
\phi^{N} \lambda^\dag \tilb =& \epsilon^{I_1...I_N}\epsilon^{i_1...i_{N+2}}\phi_{ i_1 I_1}...\phi_{i_N I_N } \lambda^\dag_{i_{N+1}i_{N+2}}\tilb
\end{aligned}
\nn\ee
where $i$ represents $SU(N+2)$ gauge indices, $a$ represents $SU(N+2)$ global indices and $I$ represents $SU(N)$ gauge indices. The first term is highly irrelevant. It does not affect the low energy physics but simply enforces the $U(1)$ symmetry structure we are looking for. The second term will be important for symmetric mass generation below.

There is a coupling constant $g^2_{N+2}$ and $g^2_{N}$ associated with each factor in the gauge group $G$. We analyse this theory in the fixed $g^2_{N+2} \gg g^2_{N}$ limit (as opposed to in \cite{us} where this ratio was varied) such that the strongly coupled dynamics of $SU(N+2)$ kick in first. We then vary the remaining parameters, the masses of the scalars $m_\phi$ and $m_\theta$, to connect the two different theories of interest.

We start by taking $m^2_\theta \gg  g_{N+2}^2 $ and $m^2_\phi \ll - g_{N+2}^2 $. $\theta$ is simply heavy and decouples from the low energy physics. We give $\phi$ the diagonal vacuum expectation value $\langle \phi_{I i}\rangle = v \delta_{I i} $. At energies $|m_\phi^2|\gg E\gg g_N$, this results exactly in the Higgs phase of the extended anti-symmetric theory presented in Section \ref{ssAS}. This is the same as the confined fermion phase of the anti-symmetric theory and leaves the massless fermions $\tchi$ in the $\sym$ of $SU(N)$ and the singlet $\rho$. All together, the following massless fermions remain:
\be
\begin{array}{|c||c||c|c|c|c|c|c|c|}
\hline & SU(N) & SU(N+2) & U(1)_1 & U(1)_2 \\ \hline
\tchi & \sym & \mathbf{1}  & 1 & 2 \\
\eta & \overline{\Box} & \Box & 0 & -1 \\
\rho  & \mathbf{1} & \mathbf{1} & N+1 & N \\ 
\tilb   & \mathbf{1} & \mathbf{1} & N+1 & N \\ \hline
\end{array}\nn\ee
with all scalars decoupling. But now the second term in \eqref{int} descends to a mass term
\be
\phi^{N} \lambda^\dag \tilb \sim v^N \rho^\dag \tilb
\nn\ee
This is the symmetric mass generation mechanism described in \cite{DTsm}. This gaps both singlet fermions and leaves us with
\be
\begin{array}{|c||c||c|c|c|c|c|c|c|}
\hline & SU(N) & SU(N+2) & U(1)_1 & U(1)_2 \\ \hline
\tchi & \sym & \mathbf{1}  & 1 & 2 \\
\eta & \overline{\Box} & \Box & 0 & -1 \\
\hline
\end{array}\nn\ee
as the IR theory. This is exactly the symmetric gauge theory of interest which at low energies flows to the coset.

Now we instead take $m^2_\phi \gg  g_{N+2}^2 $ and $m^2_\theta \ll  -g_{N+2}^2 $. $\phi$ decouples and we give $\theta$ a colour-flavour locking vacuum expectation value $\langle \theta^i_a \rangle = v \delta^i_a $ which completely Higgses the $SU(N+2)$ gauge group
\be
SU(N+2)_\text{gauge} \times SU(N+2)_\text{global} \rightarrow SU(N+2)'_\text{global}
\nn\ee
leaving the remaining global symmetries intact. From the Yukawa term $\sim \theta \tilde{\psi}\eta$ in \eqref{int}, the vacuum gaps all $\tilde{\psi}$ and $\eta$ and leaves the massless fermions
\be
\begin{array}{|c||c|c|c|c|c|c|}
\hline & SU(N+2) & U(1)_1 & U(1)_2 \\ \hline
\rule{0pt}{2.5ex} 
\lambda & \anti & 1 & 0 \\
\tilb   & \mathbf{1} & N+1 & N \\ \hline
\end{array}\nn\ee
in addition to a decoupled $SU(N)$ Yang-Mills which is gapped. The low energy physics is that of the massless confined fermions. The $U(1)$ charges can be matched to the previous basis in \eqref{symUV} by taking $Q_R=Q_1$ and $Q_L=2Q_1-Q_2$.

\section{2d Bars-Yankielowicz Models}
\label{general}
In this final section, we study chiral gauge theories which are simple generalisations of the symmetric and anti-symmetric theories considered in Section \ref{p0theories}. These extended theories have the same chiral matter as before but with $p$ additional Dirac fermions. They correspond to 2d analogues of the 4d Bars-Yankielowicz models \cite{by}.

We focus on the extension of the anti-symmetric theory. The story presented is nearly identical for the symmetric theory. Consider an $SU(N)$ gauge theory with a single right-moving Weyl fermion in the $\anti$. In addition, we have $p$ right-moving fermions in the $\overline{\Box}$ and a further $q$ left-moving fermions also in the $\overline{\Box}$. Gauge anomaly cancellation requires $q= N-2+p$ and the continuous global symmetry group is
\be
G=SU(N-2+p)\times SU(p)\times U(1)_1 \times U(1)_2 \times U(1)_2
\nn\ee
Under the combination $SU(N)\times G$, the full collection of fermions transforms as
\be
\begin{array}{|c||c||c|c|c|c|c|c|}
\hline & SU(N) & SU(N-2+p) & SU(p) & U(1)_1 & U(1)_2& U(1)_3  \\ \hline 
\rule{0pt}{2.5ex} 
\tilde{\chi} & \anti & \mathbf{1} & \mathbf{1} & 1 & 0 & 0 \\
\rule{0pt}{2.5ex} 
\psi & \overline{\Box} & \Box & \mathbf{1} & 0 & 1 & 0  \\
\tilde{\rho} & \overline{\Box}  & \mathbf{1} & \Box & 0 & 0 & 1 \\
\hline
\end{array}\label{antisym}\ee
The theory can be deformed by the four-fermion terms
\be
\begin{aligned}
\mathcal{L}_4^\text{UV} &= a_1 \operatorname{tr}(\tchi^\dag \tchi) \operatorname{tr}(\psi^\dag \psi) +a_2 \operatorname{tr}(\trho^\dag \trho) \operatorname{tr}(\psi^\dag \psi)\\
&+ a_3\, \tchi^\dag_{ij}\tchi^{jk} (\psi^\dag)^i_a \psi_k^a
+ \, a_4 (\trho^\dag)^i_I \trho_j^I (\psi^\dag)^j_a\psi_i^a
\end{aligned}
\label{as4fermi}\ee
where the $i$ and $j$ indices are associated with the $SU(N)$ gauge group, the $a$ indices with the $SU(N-2+p)$ flavour group and the $I$ indices with the $SU(p)$ global symmetry. The first two terms are again $U(1)$ current-current interactions and remain exactly marginal under RG flow. The remaining two terms are gauge current-current interactions. They mix colour indices between the different fermions and will generically pick up quantum corrections, making them marginally relevant or irrelevant.

There are two non-Abelian 't Hooft anomalies,
\be 
{\cal A}[SU(q)^2] = - {\cal A}[SU(p)^2]=N
\label{asanom1}
\ee
and a further three Abelian anomalies,
\be
\begin{aligned}
&{\cal A}[U(1)_1^2]=-\dfrac{N}{2}(N-1)\\
&{\cal A}[U(1)_2^2]= N(N-2+p)\\
&{\cal A}[U(1)_3^2]=-Np\\
\end{aligned}
\label{asanom2}
\ee
with all mixed $U(1)_a\cdot U(1)_b$ anomalies vanishing for $a\neq b$. Finally, there is the gravitational anomaly
\be
{\cal A}[\text{Grav}]=c_L-c_R= \dfrac{N}{2}(N-3)
\label{asanom3}
\ee
This theory again has two candidate IR phases.

The first is the coset proposal. Applying \eqref{QCDcoset1} gives the IR CFT
\be
\mathcal{T}=\dfrac{U (N(N-2+p))_{1}}{SU(N)_{N-2+p}} \otimes  \left[ \dfrac{U(N(N-1)/2)_1 \oplus U(Np)_{1}}{SU(N)_{N-2+p}} \right]
\nn\ee
We get the diagonal coset for the right current as the right-handed $\tchi$ and $\trho$ are in different representations of the gauge group. The branching rules of the left-current coset are again known by level-rank duality which leads to the equivalent CFT
\be
U(N-2+p)_N \otimes  \left[ \dfrac{U(N(N-1)/2)_1 \oplus U(Np)_{1}}{SU(N)_{N-2+p}} \right]
\label{eq:BYcoset}\ee
The central charge of the CFT is
\be
c_\text{coset}= \dfrac{2N(N^2-1)}{2N-2+p}-\dfrac{N(N+5)}{2}+2Np+2
\label{ascosetc}\ee

The anomalies can also be matched by the following confined, massless, gauge-singlet fermions
\be
\lambda_S = \psi \tchi \psi, \;\;\;\; \lambda_A = \tilde{\rho}(\partial_{+}\tilde{\chi})(\partial_{+}\tilde{\rho}),  \;\;\;\; \tilde{\lambda}_B = \trho \tchi \psi,\;\;\;\;  \tilde{\mathcal{B}} \sim \psi^{N-2+p}\tchi^{N-1} (\trho^\dag)^p
\nn\ee
Fermi statistics fix $\lambda_S$ to have symmetric flavour indices whereas we have chosen to anti-symmetrise the flavour indices of $\lambda_A$. The baryonic operator $\tilb$ can be constructed as
\be
\tilb = \epsilon_{a_1...a_{N-2+p}} \psi_{i_1}^{a_1}...  \psi_{i_{N-2+p}}^{a_{N-2+p}}\epsilon_{j_1...j_N} \tchi^{i_1j_1}... \tchi^{i_{N-2}j_{N-2}}\tchi^{j_{N-1}j_N}\epsilon^{I_1...I_p}(\trho^\dag)_{I_1}^{i_{N-1}}...(\trho^\dag)_{I_p}^{i_{N-2+p}}
\nn\ee
with the same index conventions as above. Under the global symmetry $G$, the quantum numbers of the fermions are
\be
\begin{array}{|c||c|c|c|c|c|c|}
\hline  & SU(N-2+p) & SU(p) & U(1)_1 & U(1)_2 & U(1)_3  \\ \hline 
\lambda_S & \sym & \mathbf{1} & 1 & 2 & 0 \\
\lambda_A & \mathbf{1} & \anti & 1 & 0 & 2  \\
\tilde{\lambda}_B & \Box  & \Box& 1 & 1 & 1 \\
\tilde{\mathcal{B}} & \mathbf{1}  & \mathbf{1} & N-1 & N-2+p & -p  \\
\hline
\end{array}\label{asfermions}\ee
It is straightforward to check that these reproduce the anomalies \eqref{asanom1}, \eqref{asanom2} and \eqref{asanom3}. The fermions again need not be free and can be deformed by the classically marginal operators
\be
\begin{aligned}
\mathcal{L}^{\text{IR}
}_{\text{4}}=& g_1\operatorname{tr}(\lambda_S^\dag \lambda_S)\operatorname{tr}(\tilde{\lambda}_B^\dag \tilde{\lambda}_B)+g_2\operatorname{tr}(\lambda_S^\dag \lambda_S)(\tilb^\dag\tilb)+g_3\operatorname{tr}(\lambda_A^\dag \lambda_A)\operatorname{tr}(\tilde{\lambda}_B^\dag \tilde{\lambda}_B)\\ +& g_4\operatorname{tr}(\lambda_A^\dag \lambda_A)(\tilb^\dag\tilb)+ g_5 (\lambda^{\dag}_{S \, ab}\lambda_{S}^{bc})(\tilde{\lambda}^{\dag}_{B \, cI} \tilde{\lambda}_{B}^{aI}) + g_6 (\lambda^{\dag}_{A \, IJ}\lambda_{A}^{JK})(\tilde{\lambda}^{\dag}_{B \, aK} \tilde{\lambda}_{B}^{aI}) \\
+& g_7\lambda_{A}^{IJ} \lambda_{S}^{ab} \tilde{\lambda}^{\dag}_{B \, aI} \tilde{\lambda}^{\dag}_{B \, bJ} +\text{h.c.}
\label{fourfermidefanti}
\end{aligned}
\ee
The first four terms are exactly marginal as they are Abelian current-current deformations. The remaining terms mix flavour between the different fermions and in general will be renormalised. We analyse this below.

The central charge of the fermions is
\be
c_\text{fermion}= \dfrac{1}{2} \left( N^2 + 4 (p-1)^2 + N (4p-3)\right)
\label{fermc}
\ee
Unlike the cases we have seen so far, where the IR proposals have been consistent with the $c$-theorem \cite{ctheorem} for all allowed $N$ and $p$, this proposal is not. The condition $c_\text{UV}>c_\text{fermion}$ requires $N(N-1) - 2 (p-1)^2>0$. This is satisfied for all $N\geq 2$ when $p=1$ but requires 
$2N \geq 1 + \sqrt{ 8 p( p-2)+9 }$ for $p\geq2$ which gives the approximately linear bound
\be
N\gtrsim \sqrt{2} p
\nn\ee
for large $N$ and $p$. For fixed $N$, the UV theory cannot flow to this fermion phase at large $p$.

\subsubsection*{Dynamics}

The central charges of the two IR candidates satisfy
\be
c_\text{fermion}>c_\text{coset}
\nn\ee
for all allowed $p$ and $N$ (except for the special case $N=2$ and $p=1$ discussed below). As the central charges are not equal, the equivalence of the two proposals and the possibility that they live on the same conformal manifold are ruled out.

So what is the low energy physics of the extended UV theory? This again depends on the four-fermi couplings $a_i$ in \eqref{as4fermi}.  We first consider the gauge theory with all $a_i=0$ and assume this flows to the coset \eqref{eq:BYcoset}. As discussed in \cite{Kiritsis}, this coset admits an interesting 't Hooft (or more precisely Veneziano-like) large N limit, $ N \rightarrow \infty$, $p \rightarrow \infty$, with
\be
\lambda  = \dfrac{p}{N}
\nn\ee
kept fixed. In the $\lambda \ll 1$ limit, $c_\text{coset} \sim \mathcal{O}(N^2)$ and the IR theory is well approximated by the $p=0$ coset \eqref{eq:BYcoset}. In the opposing $\lambda \gg 1$ limit, the central charge instead scales as $c_\text{coset} \sim \mathcal{O}(N p)$ and the IR coset is approximately $U(p)_N\times U(p)_N$. This is the coset describing the low energy dynamics of $SU(N)$ QCD with $p$ flavours, which in the $\lambda \gg 1$ limit approaches $pN$ free Dirac fermions \cite{Kiritsis,Delmastro2}.

Now turning on small four fermion deformations ($|a_i| \ll 1$) corresponds to small deformations of the coset. In the $\lambda \gg 1$ limit, depending on which UV deformations we turn on, this induces Thirring or Gross-Neveu type interactions on the approximately free $pN$ Dirac fermions.

As we make the four-fermion deformations large, it is difficult to be precise about the low energy physics. We conjecture that in the $N\gtrsim \sqrt{2} p$ region there is a finely tuned $a_i=a_i^*\gg 1$, which might be at infinity, that flows to the confined free fermions \eqref{asfermions}. Specifically, the operators in the gauge theory must be fine tuned to set the deformations of the fermions \eqref{fourfermidefanti} to zero.

\begin{figure}[h]
	\centering
	\includegraphics[width=0.40 \linewidth]{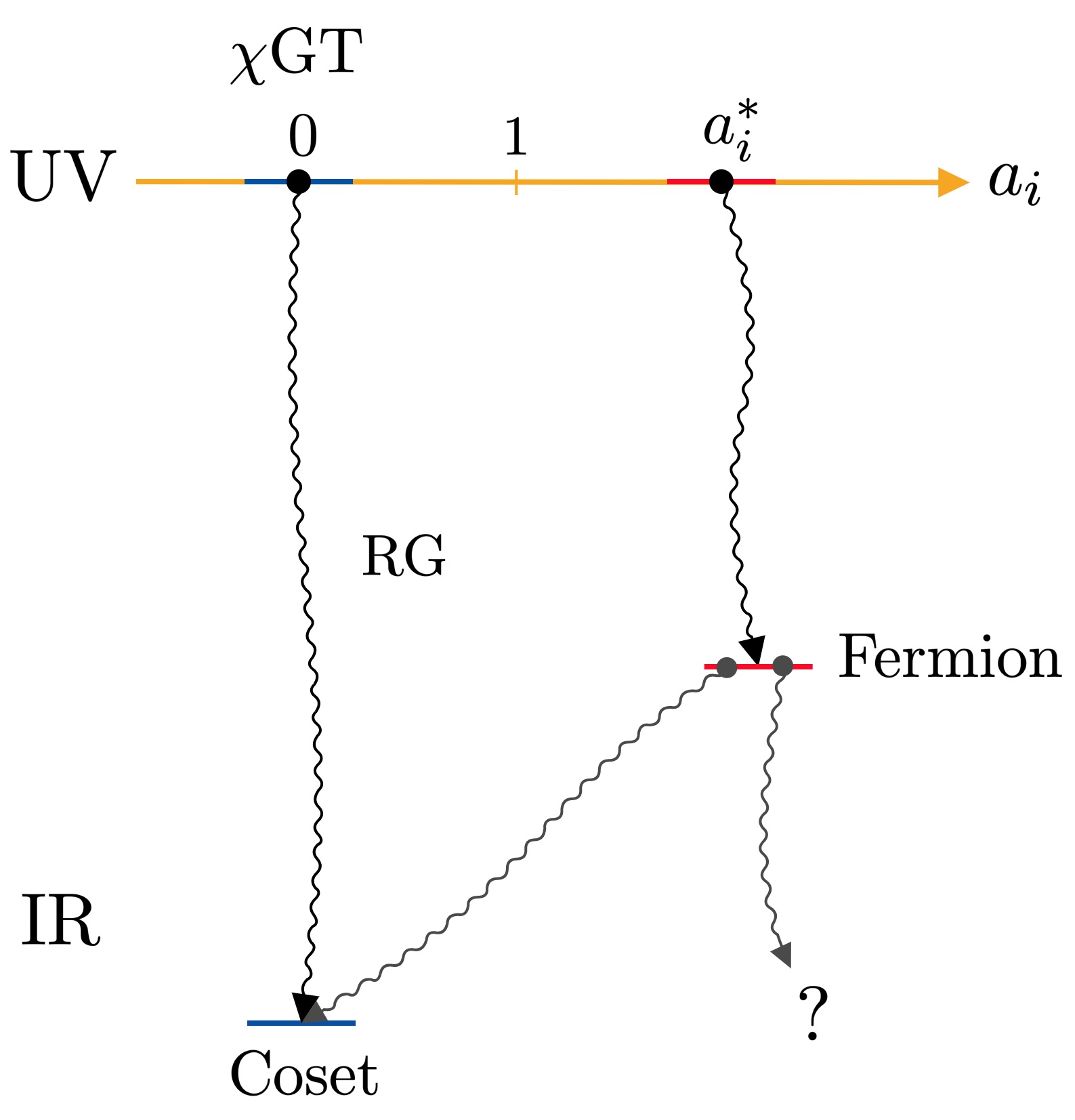}
	\caption{Proposed RG flows of the UV gauge theory ($\chi$GT), in the $N\gtrsim \sqrt{2} p$ region, as it is deformed by four-fermi interactions with couplings $a_i$. At $a_i=0$, the theory flows to the coset, and at finely tuned $a_i=a_i^*$ it flows to free fermions. Shifting slightly away from $a^*_i$ generically turns on marginally relevant four-fermi operators of the fermions, driving the theory to a new fixed point. This may be the coset or a different CFT.}
	\label{flow2}
\end{figure}

Moving slightly away from the finely tuned $a^*_i$ will turn on the four-fermion deformations of the IR theory. As we show below, some of these deformations are marginally relevant and will drive the fermions to a new fixed point. We do not know what this fixed point is. One option is that the fermions flow to the coset. This is schematically shown in Figure \ref{flow2}.

The $p=1$ and $N=2$ case is special. The UV theory has the matter content of $N_f=1$ $SU(2)$ QCD and an additional singlet fermion $\tchi$. Crucially there is no non-Abelian global symmetry and the central charges of the candidate phases match $c_\text{fermion}=c_\text{coset}$. The coset reduces to
$U(1)_2^L\otimes U(1)_2^R \otimes U(1)_1^R$, with $L$ and $R$ indicating the chiral/anti-chiral nature of the CFT. This is simply the coset of $SU(2)$ QCD with $N_f=1$ in addition to $U(1)_1^R$ which is equivalent to the decoupled free fermion $\tchi$. The fermion phase instead consists of three fermions $\lambda_S$, $\tilde{\lambda}_B$ and $\tilb$ with mixed charges under the $U(1)$'s.

We propose that the RG flows of this theory are similar to the $p=0$ case studied in Section \ref{ssAS}. We again assume the UV theory with no four-fermi deformations flows to the coset. This is of course consistent with the free fermion $\tchi$ remaining decoupled from the QCD theory. We then propose that turning on large four-fermion deformations in the UV will lead the theory to flow to free fermions. We conjecture that both IR theories live on the same conformal manifold $\tilde{\mathcal{M}}_\text{IR}$. They can be continuously connected to each other by exactly marginal $U(1)$ current-current deformations. The novelty here is that $\tilde{\mathcal{M}}_\text{IR}$ is two-dimensional and is parameterised by the current-current deformations $\sim J^L \bJ^R_1 + J^L \bJ^R_2 $ as there are two independent anti-chiral currents. What makes this particular case special is that there are no non-Abelian symmetries. This means that the classically marginal deformations of the two IR theories must remain exactly marginal, and leads to the existence of a conformal moduli space.

The same arguments we have presented in this section go through for the extended version of the symmetric theory. There are again the coset and confined massless fermion low energy phases. The flow to the free fermion theory is again prohibited for large $p$ at finite $N$ and the conjectured RG flows take the form sketched in Figure \ref{flow2}. The only difference is that the $p=1$, $N=2$ special case no longer exists. This is a consequence of the $c$-theorem ruling out the fermion phase in the IR for these values.

\subsection*{Renormalisation of the Fermion Phase}
We have seen that the free fermion phase \eqref{asfermions} admits seven four-fermion deformations
\be
\begin{aligned}
\mathcal{L}^{\text{IR}
}_{\text{4}}=& g_1\operatorname{tr}(\lambda_S^\dag \lambda_S)\operatorname{tr}(\tilde{\lambda}_B^\dag \tilde{\lambda}_B)+g_2\operatorname{tr}(\lambda_S^\dag \lambda_S)(\tilb^\dag\tilb)+g_3\operatorname{tr}(\lambda_A^\dag \lambda_A)\operatorname{tr}(\tilde{\lambda}_B^\dag \tilde{\lambda}_B)\\ +& g_4\operatorname{tr}(\lambda_A^\dag \lambda_A)(\tilb^\dag\tilb)+ g_5 (\lambda^{\dag}_{S \, ab}\lambda_{S}^{bc})(\tilde{\lambda}^{\dag}_{B \, cI} \tilde{\lambda}_{B}^{aI}) + g_6 (\lambda^{\dag}_{A \, IJ}\lambda_{A}^{JK})(\tilde{\lambda}^{\dag}_{B \, aK} \tilde{\lambda}_{B}^{aI}) \\
+& g_7\lambda_{A}^{IJ} \lambda_{S}^{ab} \tilde{\lambda}^{\dag}_{B \, aI} \tilde{\lambda}^{\dag}_{B \, bJ} +\text{h.c.}
\end{aligned}\label{ff2}
\ee
These are classically marginal. In the quantum theory, $J\bar{J}$ deformations of the $U(1)$ currents remain exactly marginal. In our case, these correspond to linear combinations of the first four terms above and lead to a  two dimensional conformal moduli space constructed out of the the $U(1)^3$ currents. The remaining operators will pick up quantum corrections, becoming marginally relevant or irrelevant. In this section, we present these beta functions to one-loop order. The details of the calculations are shown in Appendix \ref{beta}.

The chiral symmetry prevents mass terms and there are no propagator corrections. Furthermore, the first four terms in \eqref{ff2} receive no one-loop corrections. This significantly simplifies the non-zero contributions and only terms that mix flavour between different fermions pick up corrections. The beta-functions read
\be\begin{aligned}
\beta(g_5)&=-\dfrac{1}{2\pi}\left[g_1g_5+g_5^2(N+p)+g_7^2p\right]\\
\beta(g_6)&=-\dfrac{1}{2\pi}\left[g_3g_6+g_6^2p+g_7^2(N+p)\right]\\
\beta(g_7)&=-\dfrac{1}{2\pi}\left[g_1 g_7+g_3g_7+g_5g_7(N+p)+g_6g_7p \right]
\end{aligned}\nn\ee
with $\beta(g_i)=0$ for $i=1,...,4$. The diagrams that contribute to the first four beta-functions do not mix different vertices and therefore vanish. This is exclusive to one-loop and these couplings will acquire corrections at higher order, leaving only the $U(1)$ current-current operators exactly marginal.

The beta-functions admit a large set of fixed point solutions such as the obvious $g_5=g_6=g_7=0$ and any $g_1,...,g_4$. The trivial Gaussian fixed point is unstable and has relevant directions. There are also more interesting fixed point solutions such as 
\be
g_1=-(N+p)g_5, \;\;\;\; g_3= -p g_6, \;\;\;\; g_7 =0
\nn\ee
with any value of the remaining couplings. The Jacobian at this fixed point has the eigenvalues
\be
\Delta_1 =-\dfrac{\pi}{2}(N+p)g_5,\;\;\;\; \Delta_2 =- \dfrac{\pi}{2} g_6
\nn\ee
with a further five zero eigenvalues such that there are two marginal directions for positive values of $g_5$ and $g_6$.

The takeaway lesson is that the IR fermion phase admits marginally relevant deformations and is unstable. To hit the free fermion phase, the four-fermion deformations \eqref{as4fermi} of the UV gauge theory must be finely tuned. 
\appendix

\section{Appendix: No Discrete Anomalies}
\label{discrete}
Throughout the main text, we only consider continuous symmetries and omit any discrete subtleties. In this appendix, we comment on discrete quotients and show that matching perturbative 't Hooft anomalies are enough to ensure that there are no extra discrete anomaly matching conditions.

For simplicity we focus on the anti-symmetric theory of Section \ref{ssAS}. Similar arguments apply to the other theories. As discussed in Section \ref{ssAS}, the naive global symmetry group of this theory is
\be
H=SU(N-2)\times U(1)_R\times U(1)_L
\nn\ee
This captures the Lie-algebraic part of the global symmetry but may have shortcomings. First, are there any additional disconnected components? The answer is no. In 2d there is no non-Abelian theta angle so there can not be any extra discrete transformations that shift the theta angle by $2\pi \mathbb{Z}$.

Second, $H$ might not act faithfully on the fermions. To find redundant
elements, consider the most general centre transformation of $SU(N)\times SU(N-2)$ accompanied by $U(1)_R\times U(1)_L$ transformations. The action on the fermions is
\be
\begin{aligned}
\tchi & \rightarrow e^{4\pi i m/N}e^{2\pi i \alpha_r}\tchi \\
\psi& \rightarrow e^{-2\pi i m/N}e^{2\pi i k/(N-2)} e^{2\pi i 
\alpha_l} \psi
\end{aligned}
\nn\ee
where $m,k \in \mathbb{Z}$ parameterise the centre of $SU(N)$ and $SU(N-2)$ respectively and $\alpha_{r,l}$ parametrises the actions of $U(1)_{R,L}$. It is always possible to eliminate a general centre transformation by $U(1)$ transformations. For the $SU(N)$ centre transformation ($m=1$, $k=0$), we can take
\be
\alpha_r=\dfrac{N-2}{N}\;\;\; \text{and}\;\;\; \alpha_l=\dfrac{1}{N}
\nn\ee
to cancel the centre transformation modulo $2\pi$. Similarly, for an $SU(N-2)$ centre transformation ($m=0$, $k=1$) we can take
\be
\alpha_l= \dfrac{N-3}{N-2}
\nn\ee
The correct faithfully acting symmetry group is then obtained by quotienting these transformations and is
\be
G=\dfrac{SU(N-2)\times U(1)_R\times U(1)_L}{\mathbb{Z}_N\times \mathbb{Z}_{N-2}}
\nn\ee

Now one might worry that this quotient leads to extra discrete anomaly matching conditions. As argued in \cite{me}, this cannot be the case. The matching of perturbative $H$ anomalies is enough to ensure anomaly matching for the faithful symmetry group $G$. This follows by simply noticing that discrete quotient can introduce anomalies only by changing the periodicity of the theta angles for background gauge fields. But as $\Pi_0(H)=0$, theta terms cannot be generated and there are no extra anomalies for $G$.

\section{Appendix: Renormalisation of Four-Fermi Interactions}
\label{beta}

In this appendix, we describe in some detail the one-loop renormalisation of four-fermion interactions involving Weyl fermions. We start by discussing the renormalisation of four-fermion terms consisting of $K$ left-handed and a single right-handed Weyl fermion. This is the form of the exactly marginal deformations of the confined fermion phases in Section \ref{p0theories}. We show that the one-loop beta-function vanishes, as it must for $J\bJ$ deformations. The $K=1$ case agrees with the famous Thirring model \cite{Thirring}. We then reproduce the large N beta-function of the chiral Gross-Neveu model \cite{gn} by working explicitly in terms of Weyl, not Dirac, fermions. Finally, we present details of the renormalisation of the 2d Bars-Yankielowicz models discussed in Section \ref{general}.

\subsection*{$K$ Left-Handed and a Single Right-Handed Weyl Fermion}
Consider $K\geq1$ left-moving Weyl fermions $\lambda^i$ and a single right-mover $\tilb$. This admits a unique four-fermi deformation preserving the $SU(K)\times U(1)^2$ symmetry with the action
\be
S=\int d^2x\; i \lambda_{i}^\dag \partial_{-}\lambda^i+ i\tilb^\dag \partial_{+}\tilb + g (\lambda_{i}^\dag \lambda^i)(\tilb^\dag \tilb)
\nn\ee
The deformation corresponds to the $J\bJ$ deformation of the $U(1)$ symmetries and is exactly marginal. We now show that the one-loop beta-function indeed vanishes.

The propagators in momentum space appear as $\sim i/k_{\pm}$, with $k_{\pm}=k_0\pm k_1$ in light-cone coordinates. The Feynman rule for the vertex is $ig$.  We draw solid lines for $\lambda$ and dotted lines for $\tilb$.

Symmetry prohibits any mass terms. Diagrammatically, this can be seen at one-loop by looking at the propagator correction, e.g., for $\langle \tilb^\dag \tilb \rangle$,
\be
	\vcenter{\hbox{\begin{tikzpicture}[x=0.55pt,y=0.55pt,yscale=-1,xscale=1]
\draw  [line width=0.75]   (140,50) -- (145.25,50) ;
\draw [shift={(147.25,50)}, rotate = 180] [color={rgb, 255:red, 0; green, 0; blue, 0 }  ][line width=0.75]    (8.74,-2.63) .. controls (5.56,-1.12) and (2.65,-0.24) .. (0,0) .. controls (2.65,0.24) and (5.56,1.12) .. (8.74,2.63)   ;
\draw [densely dotted]  [line width=0.75]  (40,120) -- (250,120) ;
\draw  [line width=0.75]   (110,85) .. controls (110,65.67) and (125.67,50) .. (145,50) .. controls (164.33,50) and (180,65.67) .. (180,85) .. controls (180,104.33) and (164.33,120) .. (145,120) .. controls (125.67,120) and (110,104.33) .. (110,85) -- cycle ;
\draw [shift={(86.25,120)}, rotate = 180] [color={rgb, 255:red, 0; green, 0; blue, 0 }  ][line width=0.75]    (8.74,-2.63) .. controls (5.56,-1.12) and (2.65,-0.24) .. (0,0) .. controls (2.65,0.24) and (5.56,1.12) .. (8.74,2.63)   ;
\draw [shift={(213.5,120)}, rotate = 180] [color={rgb, 255:red, 0; green, 0; blue, 0 }  ][line width=0.75]    (8.74,-2.63) .. controls (5.56,-1.12) and (2.65,-0.24) .. (0,0) .. controls (2.65,0.24) and (5.56,1.12) .. (8.74,2.63)   ;
\end{tikzpicture}
}} = -(ig) K \int \dfrac{d^2k}{(2\pi)^2} \dfrac{i}{k_{+}}
\nn\ee
at zero external momentum. Wick rotating takes the propagator factor $k_{\pm}\rightarrow k_{\pm}^E=ik_0^E\pm k_1$ with $k_0=ik_0^E$. Then, the Euclidean integral takes the form
\be
(ig) K i \int \dfrac{d^2k^E}{(2\pi)^2} \dfrac{i}{k^E_{+}}
\nn
\ee
with $k^E_{+}\in \mathbb{C}$ such that the integral is over the the whole complex plane. This vanishes as a corollary of
\be
\int_\mathbb{C}d^2z\, z^{k}=0
\label{Cintegral} \ee
for all $k\in \mathbb{Z}\setminus\{0\}$.

Moving onto the vertex, the one-loop corrections are given by the diagrams
\be
\vcenter{\hbox{\begin{tikzpicture}[x=0.55pt,y=0.55pt,yscale=-1,xscale=1]
\draw  [densely dotted]  [line width=0.75]  (120,110) -- (69.88,160.25) ;
\draw [shift={(100,90)}, rotate = 225] [color={rgb, 255:red, 0; green, 0; blue, 0 }  ][line width=0.75]    (8.74,-2.63) .. controls (5.56,-1.12) and (2.65,-0.24) .. (0,0) .. controls (2.65,0.24) and (5.56,1.12) .. (8.74,2.63)   ;
\draw [shift={(94.94,135.13)}, rotate = 135.91] [color={rgb, 255:red, 0; green, 0; blue, 0 }  ][line width=0.75]    (8.74,-2.63) .. controls (5.56,-1.12) and (2.65,-0.24) .. (0,0) .. controls (2.65,0.24) and (5.56,1.12) .. (8.74,2.63)   ;
\draw  [densely dotted]  [line width=0.75]  (200.38,110) -- (250.38,160) ;
\draw  [line width=0.75]    (250.38,60) -- (200,110.5) ;
\draw [shift={(226.13,84.5)}, rotate = 134.63] [color={rgb, 255:red, 0; green, 0; blue, 0 }  ][line width=0.75]    (8.74,-2.63) .. controls (5.56,-1.12) and (2.65,-0.24) .. (0,0) .. controls (2.65,0.24) and (5.56,1.12) .. (8.74,2.63)   ;
\draw [shift={(229.63,139.25)}, rotate = 224.41] [color={rgb, 255:red, 0; green, 0; blue, 0 }  ][line width=0.75]    (8.74,-2.63) .. controls (5.56,-1.12) and (2.65,-0.24) .. (0,0) .. controls (2.65,0.24) and (5.56,1.12) .. (8.74,2.63)   ;
\draw   [line width=0.75]   (120,110) .. controls (124.3,89.38) and (138.08,78.15) .. (153.21,76.03) .. controls (172.08,73.38) and (193.05,84.89) .. (200.38,110) ;
\draw [shift={(162.25,75.75)}, rotate = 181.51] [color={rgb, 255:red, 0; green, 0; blue, 0 }  ][line width=0.75]    (8.74,-2.63) .. controls (5.56,-1.12) and (2.65,-0.24) .. (0,0) .. controls (2.65,0.24) and (5.56,1.12) .. (8.74,2.63)   ;
\draw   [line width=0.75]   (70,60) -- (120,110) ;
\draw  [densely dotted]  [line width=0.75]  (120,110) .. controls (128.5,158) and (191.25,159) .. (200,110.5) ;
\draw [shift={(163,146.5)}, rotate = 180] [color={rgb, 255:red, 0; green, 0; blue, 0 }  ][line width=0.75]    (8.74,-2.63) .. controls (5.56,-1.12) and (2.65,-0.24) .. (0,0) .. controls (2.65,0.24) and (5.56,1.12) .. (8.74,2.63)   ;
\end{tikzpicture}}}\;\;\;\;\;+\;\;\;\;\;
\vcenter{\hbox{\begin{tikzpicture}[x=0.55pt,y=0.55pt,yscale=-1,xscale=1]
\draw  [densely dotted]  [line width=0.75]  (120,110) -- (69.88,160.25) ;
\draw [shift={(100,90)}, rotate = 225] [color={rgb, 255:red, 0; green, 0; blue, 0 }  ][line width=0.75]    (8.74,-2.63) .. controls (5.56,-1.12) and (2.65,-0.24) .. (0,0) .. controls (2.65,0.24) and (5.56,1.12) .. (8.74,2.63)   ;
\draw [shift={(89,141)}, rotate = 315] [color={rgb, 255:red, 0; green, 0; blue, 0 }  ][line width=0.75]    (8.74,-2.63) .. controls (5.56,-1.12) and (2.65,-0.24) .. (0,0) .. controls (2.65,0.24) and (5.56,1.12) .. (8.74,2.63)   ;
\draw [densely dotted]  [line width=0.75]  (200.38,110) -- (250.38,160) ;
\draw  [line width=0.75]   (250.38,60) -- (200,110.5) ;
\draw [shift={(226.13,84.5)}, rotate = 134.63] [color={rgb, 255:red, 0; green, 0; blue, 0 }  ][line width=0.75]    (8.74,-2.63) .. controls (5.56,-1.12) and (2.65,-0.24) .. (0,0) .. controls (2.65,0.24) and (5.56,1.12) .. (8.74,2.63)   ;
\draw [shift={(220.63,130)}, rotate = 45.41] [color={rgb, 255:red, 0; green, 0; blue, 0 }  ][line width=0.75]    (8.74,-2.63) .. controls (5.56,-1.12) and (2.65,-0.24) .. (0,0) .. controls (2.65,0.24) and (5.56,1.12) .. (8.74,2.63)   ;
\draw   [line width=0.75]  (120,110) .. controls (124.3,89.38) and (138.08,78.15) .. (153.21,76.03) .. controls (172.08,73.38) and (193.05,84.89) .. (200.38,110) ;
\draw [shift={(162.25,75.75)}, rotate = 181.51] [color={rgb, 255:red, 0; green, 0; blue, 0 }  ][line width=0.75]    (8.74,-2.63) .. controls (5.56,-1.12) and (2.65,-0.24) .. (0,0) .. controls (2.65,0.24) and (5.56,1.12) .. (8.74,2.63)   ;
\draw   [line width=0.75]  (70,60) -- (120,110) ;
\draw [densely dotted]  [line width=0.75]  (120,110) .. controls (128.5,158) and (191.25,159) .. (200,110.5) ;
\draw [shift={(156.5,146.5)}, rotate = 358.09] [color={rgb, 255:red, 0; green, 0; blue, 0 }  ][line width=0.75]    (8.74,-2.63) .. controls (5.56,-1.12) and (2.65,-0.24) .. (0,0) .. controls (2.65,0.24) and (5.56,1.12) .. (8.74,2.63)   ;
\end{tikzpicture}}}
\nn\ee

\noindent 
The two diagrams differ by a minus sign because of the orientation of the fermion arrows and the two contributions cancel. As expected, the vertex is explicitly not renormalised to one-loop order.

It is also worth noting that Lorentz symmetry prohibits four-fermi terms such as $|\tilb|^4$. In terms of diagrams, the one loop corrections to such a vertex have the form $\int  \dfrac{d^2k}{(2\pi)^2}\dfrac{1}{k_{-}^2} $ which, after Wick rotating, vanish following \eqref{Cintegral}, as required. Any fermion loop that contains only left-moving or only right-moving fermions must necessarily vanish.

\subsection*{Chiral Gross-Neveu Model}
The chiral Gross-Neveu model consists of $N$ Dirac fermions $\Psi^i$ interacting through a four-fermi term. In the chiral basis, we write the Dirac fermions in terms of their Weyl fermions as $\Psi^i=(\rho^i, \tchi^i)^T$ such that the interaction term is
\be
\mathcal{L_{\chi\text{GN}}}=\dfrac{g^2}{2}\left[(\bar{\Psi}_i\Psi^i)^2- (\bar{\Psi}_i\gamma^5\Psi^i)^2\right]=g^2(\tchi^\dag_i \rho^i)(\rho^\dag_j \tchi^j)
\nn\ee
Crucially, the flavour indices mix between the left and right-moving fermions. We now reproduce the famous large $N$ asymptotic freedom of this model while working with Weyl fermions. This will serve as a nice warm-up for the more involved calculations below. We use the same conventions as above with solid lines for $\rho$ and dotted lines for  $\tchi$.

As above, any mass corrections vanish. The only one-loop vertex contribution at large $N$ is
\be
	\Gamma= \vcenter{\hbox{\begin{tikzpicture}[x=0.55pt,y=0.55pt,yscale=-1,xscale=1]
\draw  [densely dotted]  [line width=0.75]  (115.5,110) -- (65.38,160.25) ;
\draw [shift={(95.5,90)}, rotate = 225] [color={rgb, 255:red, 0; green, 0; blue, 0 }  ][line width=0.75]    (8.74,-2.63) .. controls (5.56,-1.12) and (2.65,-0.24) .. (0,0) .. controls (2.65,0.24) and (5.56,1.12) .. (8.74,2.63)   ;
\draw [shift={(84.5,141)}, rotate = 315] [color={rgb, 255:red, 0; green, 0; blue, 0 }  ][line width=0.75]    (8.74,-2.63) .. controls (5.56,-1.12) and (2.65,-0.24) .. (0,0) .. controls (2.65,0.24) and (5.56,1.12) .. (8.74,2.63)   ;
\draw  [densely dotted]  [line width=0.75]  (205.63,110) -- (255.63,160) ;
\draw  [line width=0.75]   (255.63,60) -- (205.25,110.5) ;
\draw [shift={(231.38,84.5)}, rotate = 134.63] [color={rgb, 255:red, 0; green, 0; blue, 0 }  ][line width=0.75]    (8.74,-2.63) .. controls (5.56,-1.12) and (2.65,-0.24) .. (0,0) .. controls (2.65,0.24) and (5.56,1.12) .. (8.74,2.63)   ;
\draw [shift={(225.88,130)}, rotate = 45.41] [color={rgb, 255:red, 0; green, 0; blue, 0 }  ][line width=0.75]    (8.74,-2.63) .. controls (5.56,-1.12) and (2.65,-0.24) .. (0,0) .. controls (2.65,0.24) and (5.56,1.12) .. (8.74,2.63)   ;
\draw  [densely dotted]  [line width=0.75]  (120,110) .. controls (124.3,89.38) and (138.08,78.15) .. (153.21,76.03) .. controls (172.08,73.38) and (193.05,84.89) .. (200.38,110) ;
\draw [shift={(155,75.5)}, rotate = 358.41] [color={rgb, 255:red, 0; green, 0; blue, 0 }  ][line width=0.75]    (8.74,-2.63) .. controls (5.56,-1.12) and (2.65,-0.24) .. (0,0) .. controls (2.65,0.24) and (5.56,1.12) .. (8.74,2.63)   ;
\draw   [line width=0.75]  (65.5,60) -- (115.5,110) ;
\draw  [line width=0.75]   (120,110) .. controls (128.5,158) and (191.25,159) .. (200,110.5) ;
\draw [shift={(162.75,146.25)}, rotate = 178.09] [color={rgb, 255:red, 0; green, 0; blue, 0 }  ][line width=0.75]    (8.74,-2.63) .. controls (5.56,-1.12) and (2.65,-0.24) .. (0,0) .. controls (2.65,0.24) and (5.56,1.12) .. (8.74,2.63)   ;
\end{tikzpicture}
}} =-(ig^2)^2 N \int \dfrac{d^2k}{(2\pi)^2}\dfrac{i}{k_+}\dfrac{i}{k_-}=-ig^4N\int \dfrac{d^2k_E}{(2\pi)^2} \dfrac{1}{-k_E^2}
\nn\ee
at vanishing external momentum and we Wick rotate for the final equality. The momentum integral can be calculated using dimensional regularisation. Working in $d=2-\epsilon$ dimensions, it reads
\be
\int \dfrac{d^dk}{(2\pi)^d}\dfrac{1}{k^2}=\dfrac{\Gamma\left(1-\dfrac{d}{2}\right)}{(4\pi)^{d/2}}=\dfrac{1}{4\pi}\left(\dfrac{2}{\epsilon}\right) + \text{regular in }\epsilon
\label{dimreg}\ee
Then writing the dimensionless coupling as $g^2=g^2(\mu)\mu^\epsilon$ results in the four-point function
\be
\Gamma \sim \dfrac{iNg^4}{4\pi}\left(\dfrac{2}{\epsilon}\right)\left( 1+\epsilon \log(\mu^2)\right)
\nn\ee
where we ignore terms regular in $\epsilon$. In the MS-scheme, we choose the counterterm $i\delta g^2=\frac{-iNg^4(\mu)}{(2\pi\epsilon)}$ to cancel the pole which leads to the effective vertex
\be
g^2_\text{eff}(\mu)=ig^2(\mu)+\dfrac{iN g^4(\mu)}{2\pi}\log(\mu^2)
\nn\ee
Finally the condition $\mu \dfrac{\partial g^2_\text{eff}}{\partial \mu}=0$ gives the beta function
\be
\beta(g^2) = \dfrac{-N (g^2)^2}{\pi}
\nn\ee
Happily, this matches the famous original result of Gross and Neveu \cite{gn}.

\subsection*{Bars-Yankielowicz Models}
Finally, we analyse, to one-loop, the low energy candidate phase of the Bars-Yankielowicz model consisting of confined fermions \eqref{asfermions}. The Feynman rules for the propagators with momentum $k_\pm$ are     
\be
\begin{aligned}
\lambda_S:& \;\; \vcenter{\hbox{\begin{tikzpicture}[x=0.7pt,y=0.7pt,yscale=-1,xscale=1]
%uncomment if require: \path (0,472); %set diagram left start at 0, and has height of 472
 \draw    (40,110) node[ above]{\scalebox{0.75}{$a$}} -- (180,110) node[above]{\scalebox{0.75}{$b$}}  ;
\draw    (40,120) node[below]{\scalebox{0.75}{$c$}} -- (180,120) node[below]{\scalebox{0.75}{$d$}};
\draw [shift={(110,110)}, rotate = 180] [color={rgb, 255:red, 0; green, 0; blue, 0 }  ][line width=0.75]    (10.93,-3.29) .. controls (6.95,-1.4) and (3.31,-0.3) .. (0,0) .. controls (3.31,0.3) and (6.95,1.4) .. (10.93,3.29)   ; 
\draw [shift={(110,120)}, rotate = 180] [color={rgb, 255:red, 0; green, 0; blue, 0 }  ][line width=0.75]    (10.93,-3.29) .. controls (6.95,-1.4) and (3.31,-0.3) .. (0,0) .. controls (3.31,0.3) and (6.95,1.4) .. (10.93,3.29)   ;
\end{tikzpicture}}}=\dfrac{i}{2k_{-}}\left(\delta^a_b \delta^c_d +\delta^a_d \delta^b_c\right)\\
\lambda_A:& \;\; \vcenter{\hbox{\begin{tikzpicture}[x=0.7pt,y=0.7pt,yscale=-1,xscale=1]
%uncomment if require: \path (0,472); %set diagram left start at 0, and has height of 472
 \draw[densely dashed]    (40,110) node[ above]{\scalebox{0.7}{$I$}} -- (180,110) node[above]{\scalebox{0.7}{$J$}}  ;
\draw[densely dashed]    (40,120) node[below]{\scalebox{0.7}{$K$}} -- (180,120) node[below]{\scalebox{0.7}{$L$}};
\draw [shift={(110,110)}, rotate = 180] [color={rgb, 255:red, 0; green, 0; blue, 0 }  ][line width=0.75]    (10.93,-3.29) .. controls (6.95,-1.4) and (3.31,-0.3) .. (0,0) .. controls (3.31,0.3) and (6.95,1.4) .. (10.93,3.29)   ; 
\draw [shift={(110,120)}, rotate = 180] [color={rgb, 255:red, 0; green, 0; blue, 0 }  ][line width=0.75]    (10.93,-3.29) .. controls (6.95,-1.4) and (3.31,-0.3) .. (0,0) .. controls (3.31,0.3) and (6.95,1.4) .. (10.93,3.29)   ;
\end{tikzpicture}}}=\dfrac{i}{2k_{-}}\left(\delta^I_J \delta^K_L -\delta^I_L \delta^J_K\right)\\
\tilde{\lambda}_B:& \;\; \vcenter{\hbox{\begin{tikzpicture}[x=0.7pt,y=0.7pt,yscale=-1,xscale=1]
%uncomment if require: \path (0,472); %set diagram left start at 0, and has height of 472
 \draw  (40,110) node[ above]{\scalebox{0.75}{$a$}} -- (180,110) node[above]{\scalebox{0.75}{$b$}}  ;
\draw[densely dashed]    (40,120) node[below]{\scalebox{0.7}{$I$}} -- (180,120) node[below]{\scalebox{0.7}{$J$}};
\draw [shift={(110,110)}, rotate = 180] [color={rgb, 255:red, 0; green, 0; blue, 0 }  ][line width=0.75]    (10.93,-3.29) .. controls (6.95,-1.4) and (3.31,-0.3) .. (0,0) .. controls (3.31,0.3) and (6.95,1.4) .. (10.93,3.29)   ; 
\draw [shift={(110,120)}, rotate = 180] [color={rgb, 255:red, 0; green, 0; blue, 0 }  ][line width=0.75]    (10.93,-3.29) .. controls (6.95,-1.4) and (3.31,-0.3) .. (0,0) .. controls (3.31,0.3) and (6.95,1.4) .. (10.93,3.29)   ;
\end{tikzpicture}}}=\dfrac{i}{k_{+}}\delta^a_b\delta^I_J \\
\tilb:& \;\;\; \;\vcenter{\hbox{\begin{tikzpicture}[x=0.7pt,y=0.7pt,yscale=-1,xscale=1]
%uncomment if require: \path (0,472); %set diagram left start at 0, and has height of 472
 \draw[line width=0.25mm,densely dotted] (40,100)  -- (180,100)   ;
\draw [shift={(110,100)}, rotate = 180] [color={rgb, 255:red, 0; green, 0; blue, 0 }  ][line width=0.75]    (10.93,-3.29) .. controls (6.95,-1.4) and (3.31,-0.3) .. (0,0) .. controls (3.31,0.3) and (6.95,1.4) .. (10.93,3.29)   ; 
\end{tikzpicture}}}\;\;=\dfrac{i}{k_{+}}\\
\end{aligned}\nn\ee
where solid lines correspond to $SU(N-2+p)$ indices and dashed lines to $SU(p)$. We indicate the symmetry structure of the first three fermions by drawing double lines which are \linebreak(anti)symmetrised where necessary. There are seven interaction terms 
\be
\begin{aligned}
\mathcal{L}^{\text{IR}
}_{\text{4}}=& g_1\operatorname{tr}(\lambda_S^\dag \lambda_S)\operatorname{tr}(\tilde{\lambda}_B^\dag \tilde{\lambda}_B)+g_2\operatorname{tr}(\lambda_S^\dag \lambda_S)(\tilb^\dag\tilb)+g_3\operatorname{tr}(\lambda_A^\dag \lambda_A)\operatorname{tr}(\tilde{\lambda}_B^\dag \tilde{\lambda}_B)\\ +& g_4\operatorname{tr}(\lambda_A^\dag \lambda_A)(\tilb^\dag\tilb)+ g_5 (\lambda^{\dag}_{S \, ab}\lambda_{S}^{bc})(\tilde{\lambda}^{\dag}_{B \, cI} \tilde{\lambda}_{B}^{aI}) + g_6 (\lambda^{\dag}_{A \, IJ}\lambda_{A}^{JK})(\tilde{\lambda}^{\dag}_{B \, aK} \tilde{\lambda}_{B}^{aI}) \\
+& g_7\lambda_{A}^{IJ} \lambda_{S}^{ab} \tilde{\lambda}^{\dag}_{B \, aI} \tilde{\lambda}^{\dag}_{B \, bJ} +\text{h.c.}
\nn
\end{aligned}
\ee
with the corresponding vertices
\be
\begin{aligned}
&\vcenter{\hbox{\begin{tikzpicture}[x=0.55pt,y=0.55pt,yscale=-1,xscale=1]
\draw    (60,60) -- (110,110) ;
\draw    (110,110) -- (60,160) ;
\draw    (60,50) -- (120,110)  ;
\draw    (120,110) -- (60,170) ;
\draw    (125,110) -- (185,170) ;
\draw[densely dashed]    (135,110) -- (185,160) ;
\draw[densely dashed]    (185,60) -- (135,110) node[right=0.3cm]{$=ig_1$} ;
\draw  (185,50) -- (125,110) ;
\draw [shift={(95,85)}, rotate = 225] [color={rgb, 255:red, 0; green, 0; blue, 0 }  ][line width=0.75]    (8.74,-2.63) .. controls (5.56,-1.12) and (2.65,-0.24) .. (0,0) .. controls (2.65,0.24) and (5.56,1.12) .. (8.74,2.63)   ;
\draw [shift={(90,90)}, rotate = 225] [color={rgb, 255:red, 0; green, 0; blue, 0 }  ][line width=0.75]    (8.74,-2.63) .. controls (5.56,-1.12) and (2.65,-0.24) .. (0,0) .. controls (2.65,0.24) and (5.56,1.12) .. (8.74,2.63)   ;
\draw [shift={(80,140)}, rotate = 315] [color={rgb, 255:red, 0; green, 0; blue, 0 }  ][line width=0.75]    (8.74,-2.63) .. controls (5.56,-1.12) and (2.65,-0.24) .. (0,0) .. controls (2.65,0.24) and (5.56,1.12) .. (8.74,2.63)   ;
\draw [shift={(85,145)}, rotate = 315] [color={rgb, 255:red, 0; green, 0; blue, 0 }  ][line width=0.75]    (8.74,-2.63) .. controls (5.56,-1.12) and (2.65,-0.24) .. (0,0) .. controls (2.65,0.24) and (5.56,1.12) .. (8.74,2.63)   ;
\draw [shift={(160,145)}, rotate = 225] [color={rgb, 255:red, 0; green, 0; blue, 0 }  ][line width=0.75]    (8.74,-2.63) .. controls (5.56,-1.12) and (2.65,-0.24) .. (0,0) .. controls (2.65,0.24) and (5.56,1.12) .. (8.74,2.63)   ;
\draw [shift={(165,140)}, rotate = 225] [color={rgb, 255:red, 0; green, 0; blue, 0 }  ][line width=0.75]    (8.74,-2.63) .. controls (5.56,-1.12) and (2.65,-0.24) .. (0,0) .. controls (2.65,0.24) and (5.56,1.12) .. (8.74,2.63)   ;
\draw [shift={(155,90)}, rotate = 315] [color={rgb, 255:red, 0; green, 0; blue, 0 }  ][line width=0.75]    (8.74,-2.63) .. controls (5.56,-1.12) and (2.65,-0.24) .. (0,0) .. controls (2.65,0.24) and (5.56,1.12) .. (8.74,2.63)   ;
\draw [shift={(150,85)}, rotate = 315] [color={rgb, 255:red, 0; green, 0; blue, 0 }  ][line width=0.75]    (8.74,-2.63) .. controls (5.56,-1.12) and (2.65,-0.24) .. (0,0) .. controls (2.65,0.24) and (5.56,1.12) .. (8.74,2.63)   ;
\end{tikzpicture}
}}, \;\; \vcenter{\hbox{\begin{tikzpicture}[x=0.55pt,y=0.55pt,yscale=-1,xscale=1]
\draw    (60,60) -- (110,110) ;
\draw    (110,110) -- (60,160) ;
\draw    (60,50) -- (120,110) node[right=0.3cm]{$=ig_2$} ;
\draw    (120,110) -- (60,170);
\draw[line width=0.25mm,densely dotted]    (125,110) -- (185,170)  ;
\draw[line width=0.25mm,densely dotted]  (185,50) -- (125,110) ;
\draw [shift={(95,85)}, rotate = 225] [color={rgb, 255:red, 0; green, 0; blue, 0 }  ][line width=0.75]    (8.74,-2.63) .. controls (5.56,-1.12) and (2.65,-0.24) .. (0,0) .. controls (2.65,0.24) and (5.56,1.12) .. (8.74,2.63)   ;
\draw [shift={(90,90)}, rotate = 225] [color={rgb, 255:red, 0; green, 0; blue, 0 }  ][line width=0.75]    (8.74,-2.63) .. controls (5.56,-1.12) and (2.65,-0.24) .. (0,0) .. controls (2.65,0.24) and (5.56,1.12) .. (8.74,2.63)   ;
\draw [shift={(80,140)}, rotate = 315] [color={rgb, 255:red, 0; green, 0; blue, 0 }  ][line width=0.75]    (8.74,-2.63) .. controls (5.56,-1.12) and (2.65,-0.24) .. (0,0) .. controls (2.65,0.24) and (5.56,1.12) .. (8.74,2.63)   ;
\draw [shift={(85,145)}, rotate = 315] [color={rgb, 255:red, 0; green, 0; blue, 0 }  ][line width=0.75]    (8.74,-2.63) .. controls (5.56,-1.12) and (2.65,-0.24) .. (0,0) .. controls (2.65,0.24) and (5.56,1.12) .. (8.74,2.63)   ;
\draw [shift={(160,145)}, rotate = 225] [color={rgb, 255:red, 0; green, 0; blue, 0 }  ][line width=0.75]    (8.74,-2.63) .. controls (5.56,-1.12) and (2.65,-0.24) .. (0,0) .. controls (2.65,0.24) and (5.56,1.12) .. (8.74,2.63)   ;
\draw [shift={(150,85)}, rotate = 315] [color={rgb, 255:red, 0; green, 0; blue, 0 }  ][line width=0.75]    (8.74,-2.63) .. controls (5.56,-1.12) and (2.65,-0.24) .. (0,0) .. controls (2.65,0.24) and (5.56,1.12) .. (8.74,2.63)   ;
\end{tikzpicture}
}},\;\; \vcenter{\hbox{\begin{tikzpicture}[x=0.55pt,y=0.55pt,yscale=-1,xscale=1]
\draw[densely dashed]     (60,60) -- (110,110) ;
\draw[densely dashed]    (110,110) -- (60,160) ;
\draw[densely dashed]     (60,50) -- (120,110)  ;
\draw[densely dashed]     (120,110) -- (60,170) ;
\draw    (125,110) -- (185,170) ;
\draw[densely dashed]    (135,110) -- (185,160) ;
\draw[densely dashed]    (185,60) -- (135,110) node[right=0.3cm]{$=ig_3$} ;
\draw  (185,50) -- (125,110) ;
\draw [shift={(95,85)}, rotate = 225] [color={rgb, 255:red, 0; green, 0; blue, 0 }  ][line width=0.75]    (8.74,-2.63) .. controls (5.56,-1.12) and (2.65,-0.24) .. (0,0) .. controls (2.65,0.24) and (5.56,1.12) .. (8.74,2.63)   ;
\draw [shift={(90,90)}, rotate = 225] [color={rgb, 255:red, 0; green, 0; blue, 0 }  ][line width=0.75]    (8.74,-2.63) .. controls (5.56,-1.12) and (2.65,-0.24) .. (0,0) .. controls (2.65,0.24) and (5.56,1.12) .. (8.74,2.63)   ;
\draw [shift={(80,140)}, rotate = 315] [color={rgb, 255:red, 0; green, 0; blue, 0 }  ][line width=0.75]    (8.74,-2.63) .. controls (5.56,-1.12) and (2.65,-0.24) .. (0,0) .. controls (2.65,0.24) and (5.56,1.12) .. (8.74,2.63)   ;
\draw [shift={(85,145)}, rotate = 315] [color={rgb, 255:red, 0; green, 0; blue, 0 }  ][line width=0.75]    (8.74,-2.63) .. controls (5.56,-1.12) and (2.65,-0.24) .. (0,0) .. controls (2.65,0.24) and (5.56,1.12) .. (8.74,2.63)   ;
\draw [shift={(160,145)}, rotate = 225] [color={rgb, 255:red, 0; green, 0; blue, 0 }  ][line width=0.75]    (8.74,-2.63) .. controls (5.56,-1.12) and (2.65,-0.24) .. (0,0) .. controls (2.65,0.24) and (5.56,1.12) .. (8.74,2.63)   ;
\draw [shift={(165,140)}, rotate = 225] [color={rgb, 255:red, 0; green, 0; blue, 0 }  ][line width=0.75]    (8.74,-2.63) .. controls (5.56,-1.12) and (2.65,-0.24) .. (0,0) .. controls (2.65,0.24) and (5.56,1.12) .. (8.74,2.63)   ;
\draw [shift={(155,90)}, rotate = 315] [color={rgb, 255:red, 0; green, 0; blue, 0 }  ][line width=0.75]    (8.74,-2.63) .. controls (5.56,-1.12) and (2.65,-0.24) .. (0,0) .. controls (2.65,0.24) and (5.56,1.12) .. (8.74,2.63)   ;
\draw [shift={(150,85)}, rotate = 315] [color={rgb, 255:red, 0; green, 0; blue, 0 }  ][line width=0.75]    (8.74,-2.63) .. controls (5.56,-1.12) and (2.65,-0.24) .. (0,0) .. controls (2.65,0.24) and (5.56,1.12) .. (8.74,2.63)   ;
\end{tikzpicture}}}, \;\;
\vcenter{\hbox{\begin{tikzpicture}[x=0.55pt,y=0.55pt,yscale=-1,xscale=1]
\draw[densely dashed]    (60,60) -- (110,110) ;
\draw[densely dashed]     (110,110) -- (60,160) ;
\draw[densely dashed]     (60,50) -- (120,110) node[right=0.3cm]{$=ig_4$} ;
\draw[densely dashed]     (120,110) -- (60,170);
\draw[line width=0.25mm,densely dotted]    (125,110) -- (185,170)  ;
\draw[line width=0.25mm,densely dotted]  (185,50) -- (125,110) ;
\draw [shift={(95,85)}, rotate = 225] [color={rgb, 255:red, 0; green, 0; blue, 0 }  ][line width=0.75]    (8.74,-2.63) .. controls (5.56,-1.12) and (2.65,-0.24) .. (0,0) .. controls (2.65,0.24) and (5.56,1.12) .. (8.74,2.63)   ;
\draw [shift={(90,90)}, rotate = 225] [color={rgb, 255:red, 0; green, 0; blue, 0 }  ][line width=0.75]    (8.74,-2.63) .. controls (5.56,-1.12) and (2.65,-0.24) .. (0,0) .. controls (2.65,0.24) and (5.56,1.12) .. (8.74,2.63)   ;
\draw [shift={(80,140)}, rotate = 315] [color={rgb, 255:red, 0; green, 0; blue, 0 }  ][line width=0.75]    (8.74,-2.63) .. controls (5.56,-1.12) and (2.65,-0.24) .. (0,0) .. controls (2.65,0.24) and (5.56,1.12) .. (8.74,2.63)   ;
\draw [shift={(85,145)}, rotate = 315] [color={rgb, 255:red, 0; green, 0; blue, 0 }  ][line width=0.75]    (8.74,-2.63) .. controls (5.56,-1.12) and (2.65,-0.24) .. (0,0) .. controls (2.65,0.24) and (5.56,1.12) .. (8.74,2.63)   ;
\draw [shift={(160,145)}, rotate = 225] [color={rgb, 255:red, 0; green, 0; blue, 0 }  ][line width=0.75]    (8.74,-2.63) .. controls (5.56,-1.12) and (2.65,-0.24) .. (0,0) .. controls (2.65,0.24) and (5.56,1.12) .. (8.74,2.63)   ;
\draw [shift={(150,85)}, rotate = 315] [color={rgb, 255:red, 0; green, 0; blue, 0 }  ][line width=0.75]    (8.74,-2.63) .. controls (5.56,-1.12) and (2.65,-0.24) .. (0,0) .. controls (2.65,0.24) and (5.56,1.12) .. (8.74,2.63)   ;\end{tikzpicture}}},\\
&\;\;\;\;\;\;\;\;\;\;\;\;\;\;\vcenter{\hbox{\begin{tikzpicture}[x=0.62pt,y=0.55pt,yscale=-1,xscale=1]
\draw    (60,60) -- (109.88,110) ;
\draw    (110.13,109.75) -- (60,160) ;
\draw    (59.5,50) -- (98.75,89.75) -- (115,106) ;
\draw    (115,114) -- (59.5,170) ;
\draw    (90,80.75) -- (93.59,84.34) ;
\draw    (114.63,113.75) -- (170.5,170) ;
\draw    (120,110) -- (170,160) [densely dashed] ;
\draw[densely dashed]     (170,60) -- (119.63,110.5) node[right=0.3cm]{$=ig_5$} ;
\draw    (170.25,49.5) -- (114.38,106) ;
\draw [shift={(95,85.75)}, rotate = 225] [color={rgb, 255:red, 0; green, 0; blue, 0 }  ][line width=0.75]    (8.74,-2.63) .. controls (5.56,-1.12) and (2.65,-0.24) .. (0,0) .. controls (2.65,0.24) and (5.56,1.12) .. (8.74,2.63)   ;
\draw [shift={(90,90)}, rotate = 225] [color={rgb, 255:red, 0; green, 0; blue, 0 }  ][line width=0.75]    (8.74,-2.63) .. controls (5.56,-1.12) and (2.65,-0.24) .. (0,0) .. controls (2.65,0.24) and (5.56,1.12) .. (8.74,2.63)   ;
\draw [shift={(80,140)}, rotate = 315] [color={rgb, 255:red, 0; green, 0; blue, 0 }  ][line width=0.75]    (8.74,-2.63) .. controls (5.56,-1.12) and (2.65,-0.24) .. (0,0) .. controls (2.65,0.24) and (5.56,1.12) .. (8.74,2.63)   ;
\draw [shift={(84.75,144.5)}, rotate = 315] [color={rgb, 255:red, 0; green, 0; blue, 0 }  ][line width=0.75]    (8.74,-2.63) .. controls (5.56,-1.12) and (2.65,-0.24) .. (0,0) .. controls (2.65,0.24) and (5.56,1.12) .. (8.74,2.63)   ;
\draw [shift={(135.5,134.75)}, rotate = 45.41] [color={rgb, 255:red, 0; green, 0; blue, 0 }  ][line width=0.75]    (8.74,-2.63) .. controls (5.56,-1.12) and (2.65,-0.24) .. (0,0) .. controls (2.65,0.24) and (5.56,1.12) .. (8.74,2.63)   ;
\draw [shift={(145.75,84.5)}, rotate = 134.63] [color={rgb, 255:red, 0; green, 0; blue, 0 }  ][line width=0.75]    (8.74,-2.63) .. controls (5.56,-1.12) and (2.65,-0.24) .. (0,0) .. controls (2.65,0.24) and (5.56,1.12) .. (8.74,2.63)   ;
\draw [shift={(140.5,79.75)}, rotate = 134.58] [color={rgb, 255:red, 0; green, 0; blue, 0 }  ][line width=0.75]    (8.74,-2.63) .. controls (5.56,-1.12) and (2.65,-0.24) .. (0,0) .. controls (2.65,0.24) and (5.56,1.12) .. (8.74,2.63)   ;
\draw [shift={(140.25,130)}, rotate = 45.41] [color={rgb, 255:red, 0; green, 0; blue, 0 }  ][line width=0.75]    (8.74,-2.63) .. controls (5.56,-1.12) and (2.65,-0.24) .. (0,0) .. controls (2.65,0.24) and (5.56,1.12) .. (8.74,2.63)   ;
\end{tikzpicture}}},\;\;\;\; \;
\vcenter{\hbox{\begin{tikzpicture}[x=0.62pt,y=0.55pt,yscale=-1,xscale=1]
\draw[densely dashed]      (60,60) -- (109.88,110) ;
\draw[densely dashed]      (110.13,109.75) -- (60,160) ;
\draw[densely dashed]      (59.5,50) -- (98.75,89.75) -- (115,106) ;
\draw[densely dashed]      (115,114) -- (59.5,170) ;
\draw[densely dashed]      (90,80.75) -- (93.59,84.34) ;
\draw[densely dashed]     (114.63,113.75) -- (170.5,170) ;
\draw    (120,110) -- (170,160)  ;
\draw    (170,60) -- (119.63,110.5) node[right=0.3cm]{$=ig_6$} ;
\draw[densely dashed]    (170.25,49.5) -- (114.38,106) ;
\draw [shift={(95,85.75)}, rotate = 225] [color={rgb, 255:red, 0; green, 0; blue, 0 }  ][line width=0.75]    (8.74,-2.63) .. controls (5.56,-1.12) and (2.65,-0.24) .. (0,0) .. controls (2.65,0.24) and (5.56,1.12) .. (8.74,2.63)   ;
\draw [shift={(90,90)}, rotate = 225] [color={rgb, 255:red, 0; green, 0; blue, 0 }  ][line width=0.75]    (8.74,-2.63) .. controls (5.56,-1.12) and (2.65,-0.24) .. (0,0) .. controls (2.65,0.24) and (5.56,1.12) .. (8.74,2.63)   ;
\draw [shift={(80,140)}, rotate = 315] [color={rgb, 255:red, 0; green, 0; blue, 0 }  ][line width=0.75]    (8.74,-2.63) .. controls (5.56,-1.12) and (2.65,-0.24) .. (0,0) .. controls (2.65,0.24) and (5.56,1.12) .. (8.74,2.63)   ;
\draw [shift={(84.75,144.5)}, rotate = 315] [color={rgb, 255:red, 0; green, 0; blue, 0 }  ][line width=0.75]    (8.74,-2.63) .. controls (5.56,-1.12) and (2.65,-0.24) .. (0,0) .. controls (2.65,0.24) and (5.56,1.12) .. (8.74,2.63)   ;
\draw [shift={(135.5,134.75)}, rotate = 45.41] [color={rgb, 255:red, 0; green, 0; blue, 0 }  ][line width=0.75]    (8.74,-2.63) .. controls (5.56,-1.12) and (2.65,-0.24) .. (0,0) .. controls (2.65,0.24) and (5.56,1.12) .. (8.74,2.63)   ;
\draw [shift={(145.75,84.5)}, rotate = 134.63] [color={rgb, 255:red, 0; green, 0; blue, 0 }  ][line width=0.75]    (8.74,-2.63) .. controls (5.56,-1.12) and (2.65,-0.24) .. (0,0) .. controls (2.65,0.24) and (5.56,1.12) .. (8.74,2.63)   ;
\draw [shift={(140.5,79.75)}, rotate = 134.58] [color={rgb, 255:red, 0; green, 0; blue, 0 }  ][line width=0.75]    (8.74,-2.63) .. controls (5.56,-1.12) and (2.65,-0.24) .. (0,0) .. controls (2.65,0.24) and (5.56,1.12) .. (8.74,2.63)   ;
\draw [shift={(140.25,130)}, rotate = 45.41] [color={rgb, 255:red, 0; green, 0; blue, 0 }  ][line width=0.75]    (8.74,-2.63) .. controls (5.56,-1.12) and (2.65,-0.24) .. (0,0) .. controls (2.65,0.24) and (5.56,1.12) .. (8.74,2.63)   ;
\end{tikzpicture}}},\;\;
\vcenter{\hbox{\begin{tikzpicture}[x=0.62pt,y=0.55pt,yscale=-1,xscale=1]
\draw    (60,60) -- (109.88,110) ;
\draw    (110.13,109.75) -- (60,160) ;
\draw    (59.5,50) -- (98.75,89.75) -- (115,106) ;
\draw[densely dashed]      (115,114) -- (59.5,170) ;
\draw[densely dashed]     (114.63,113.75) -- (170.5,170) ;
\draw[densely dashed]   (120,110) -- (170,160)  ;
\draw[densely dashed]  (170,60) -- (119.63,110.5) node[right=0.3cm]{$=ig_7$} ;
\draw    (170.25,49.5) -- (114.38,106) ;
\draw [shift={(95,85.75)}, rotate = 225] [color={rgb, 255:red, 0; green, 0; blue, 0 }  ][line width=0.75]    (8.74,-2.63) .. controls (5.56,-1.12) and (2.65,-0.24) .. (0,0) .. controls (2.65,0.24) and (5.56,1.12) .. (8.74,2.63)   ;
\draw [shift={(90,90)}, rotate = 225] [color={rgb, 255:red, 0; green, 0; blue, 0 }  ][line width=0.75]    (8.74,-2.63) .. controls (5.56,-1.12) and (2.65,-0.24) .. (0,0) .. controls (2.65,0.24) and (5.56,1.12) .. (8.74,2.63)   ;
\draw [shift={(80,140)}, rotate = 315] [color={rgb, 255:red, 0; green, 0; blue, 0 }  ][line width=0.75]    (8.74,-2.63) .. controls (5.56,-1.12) and (2.65,-0.24) .. (0,0) .. controls (2.65,0.24) and (5.56,1.12) .. (8.74,2.63)   ;
\draw [shift={(84.75,144.5)}, rotate = 315] [color={rgb, 255:red, 0; green, 0; blue, 0 }  ][line width=0.75]    (8.74,-2.63) .. controls (5.56,-1.12) and (2.65,-0.24) .. (0,0) .. controls (2.65,0.24) and (5.56,1.12) .. (8.74,2.63)   ;
\draw [shift={(135.5,134.75)}, rotate = 45.41] [color={rgb, 255:red, 0; green, 0; blue, 0 }  ][line width=0.75]    (8.74,-2.63) .. controls (5.56,-1.12) and (2.65,-0.24) .. (0,0) .. controls (2.65,0.24) and (5.56,1.12) .. (8.74,2.63)   ;
\draw [shift={(145.75,84.5)}, rotate = 134.63] [color={rgb, 255:red, 0; green, 0; blue, 0 }  ][line width=0.75]    (8.74,-2.63) .. controls (5.56,-1.12) and (2.65,-0.24) .. (0,0) .. controls (2.65,0.24) and (5.56,1.12) .. (8.74,2.63)   ;
\draw [shift={(140.5,79.75)}, rotate = 134.58] [color={rgb, 255:red, 0; green, 0; blue, 0 }  ][line width=0.75]    (8.74,-2.63) .. controls (5.56,-1.12) and (2.65,-0.24) .. (0,0) .. controls (2.65,0.24) and (5.56,1.12) .. (8.74,2.63)   ;
\draw [shift={(140.25,130)}, rotate = 45.41] [color={rgb, 255:red, 0; green, 0; blue, 0 }  ][line width=0.75]    (8.74,-2.63) .. controls (5.56,-1.12) and (2.65,-0.24) .. (0,0) .. controls (2.65,0.24) and (5.56,1.12) .. (8.74,2.63)   ;
\end{tikzpicture}}}
\end{aligned}\nn\ee
We can now compute the one-loop corrections. There are a number of observations that significantly simplify the calculation. First, as we saw for the $p=0$ case above, there are no propagator corrections. Second, any fermion loop containing a fermion of single chirality (i.e. a loop with only a left-moving fermion) is identically zero as a corollary of \eqref{Cintegral}. Finally, any vertex correction to interactions of the form $\operatorname{tr}(\text{LH}\times \text{LH})\operatorname{tr}(\text{RH}\times \text{RH})$, for left/right-handed fermions with the trace over the flavour indices, is zero. This is due to a similar cancellation as in the $p=0$ case above where contributions from two diagrams of opposing internal fermion arrows exactly cancel. This means that $g_1,...,g_4$ receive no one-loop corrections and only operators that mix flavour get modified.

Nontrivial corrections are due to diagrams that mix flavour, such as
\be\begin{aligned}
&\vcenter{\hbox{\begin{tikzpicture}[x=0.50pt,y=0.50pt,yscale=-1,xscale=1]
\draw    (60,60) -- (109.88,110) ;
\draw    (110.13,109.75) -- (60,160) ;
\draw[densely dashed]     (59.5,50) -- (98.75,89.75) -- (115,106) ;
\draw    (115,114) -- (59.5,170) ;
\draw  [shift={(95,85.75)}, rotate = 225] (8.74,-2.63) .. controls (5.56,-1.12) and (2.65,-0.24) .. (0,0) .. controls (2.65,0.24) and (5.56,1.12) .. (8.74,2.63)   ;
\draw  [shift={(90,90)}, rotate = 225]    (8.74,-2.63) .. controls (5.56,-1.12) and (2.65,-0.24) .. (0,0) .. controls (2.65,0.24) and (5.56,1.12) .. (8.74,2.63)   ;
\draw [shift={(80,140)}, rotate = 315]  (8.74,-2.63) .. controls (5.56,-1.12) and (2.65,-0.24) .. (0,0) .. controls (2.65,0.24) and (5.56,1.12) .. (8.74,2.63)   ;
\draw [shift={(84.75,144.5)}, rotate = 315]   (8.74,-2.63) .. controls (5.56,-1.12) and (2.65,-0.24) .. (0,0) .. controls (2.65,0.24) and (5.56,1.12) .. (8.74,2.63)   ;
\draw    (204.13,113.75) -- (260,170) ;
\draw [shift={(225,134.75)}, rotate = 45.41] (8.74,-2.63) .. controls (5.56,-1.12) and (2.65,-0.24) .. (0,0) .. controls (2.65,0.24) and (5.56,1.12) .. (8.74,2.63)   ;
\draw    (209.5,110) -- (259.5,160) ;
\draw   (259.5,60) -- (209.13,110.5) ;
\draw [shift={(235.25,84.5)}, rotate = 134.63](8.74,-2.63) .. controls (5.56,-1.12) and (2.65,-0.24) .. (0,0) .. controls (2.65,0.24) and (5.56,1.12) .. (8.74,2.63)   ;
\draw [densely dashed]    (259.75,49.5) -- (203.88,106) ;
\draw [shift={(230,79.75)}, rotate = 134.58] (8.74,-2.63) .. controls (5.56,-1.12) and (2.65,-0.24) .. (0,0) .. controls (2.65,0.24) and (5.56,1.12) .. (8.74,2.63)   ;
\draw [shift={(229.75,130)}, rotate = 45.41] (8.74,-2.63) .. controls (5.56,-1.12) and (2.65,-0.24) .. (0,0) .. controls (2.65,0.24) and (5.56,1.12) .. (8.74,2.63)   ;
\draw [densely dashed]    (115,106) .. controls (127.88,57.75) and (187.88,58.5) .. (203.88,106) ;
\draw    (115,114) .. controls (130.88,160.75) and (186.13,162.5) .. (203.88,114) ;
\draw    (119.88,110) .. controls (132.88,64) and (183.38,67.25) .. (197.88,110.25) ; 
\draw    (119.88,110) .. controls (136.38,153) and (180.38,155) .. (197.88,110.25) ;
\draw [shift={(160.5,70.25)}, rotate = 181.51](8.74,-2.63) .. controls (5.56,-1.12) and (2.65,-0.24) .. (0,0) .. controls (2.65,0.24) and (5.56,1.12) .. (8.74,2.63)   ;
\draw [shift={(160.75,76.75)}, rotate = 181.51]  (8.74,-2.63) .. controls (5.56,-1.12) and (2.65,-0.24) .. (0,0) .. controls (2.65,0.24) and (5.56,1.12) .. (8.74,2.63)   ;
\draw [shift={(156,143)}, rotate = 355.91](8.74,-2.63) .. controls (5.56,-1.12) and (2.65,-0.24) .. (0,0) .. controls (2.65,0.24) and (5.56,1.12) .. (8.74,2.63)   ;
\draw [shift={(156.5,149.75)}, rotate = 355.91] (8.74,-2.63) .. controls (5.56,-1.12) and (2.65,-0.24) .. (0,0) .. controls (2.65,0.24) and (5.56,1.12) .. (8.74,2.63)   ;
\end{tikzpicture}}} = \dfrac{(q+1)}{2}g_5^2 I
\end{aligned}
\nn\ee
where we write $q=N-2+p$ and define the Wick-rotated integral
\be
I=i\int \dfrac{d^2k}{(2\pi)^2}\dfrac{1}{k^2}
\nn\ee
which is common to all such diagrams at zero external momentum. The numerical factor in front is due to the fermion loop and the symmetry of $\lambda_S$. There are the further diagrams
\begingroup
\allowdisplaybreaks
\begin{align*}
&\vcenter{\hbox{\begin{tikzpicture}[x=0.5pt,y=0.5pt,yscale=-1,xscale=1]
\draw    (60,60) -- (109.88,110) ;
\draw    (110.13,109.75) -- (60,160) ;
\draw[densely dashed]     (59.5,50) -- (98.75,89.75) -- (115,106) ;
\draw    (115,114) -- (59.5,170) ;
\draw  [shift={(95,85.75)}, rotate = 225]  (8.74,-2.63) .. controls (5.56,-1.12) and (2.65,-0.24) .. (0,0) .. controls (2.65,0.24) and (5.56,1.12) .. (8.74,2.63)   ;
\draw  [shift={(90,90)}, rotate = 225]  (8.74,-2.63) .. controls (5.56,-1.12) and (2.65,-0.24) .. (0,0) .. controls (2.65,0.24) and (5.56,1.12) .. (8.74,2.63)   ;
\draw [shift={(80,140)}, rotate = 315]   (8.74,-2.63) .. controls (5.56,-1.12) and (2.65,-0.24) .. (0,0) .. controls (2.65,0.24) and (5.56,1.12) .. (8.74,2.63)   ;
\draw [shift={(84.75,144.5)}, rotate = 315] (8.74,-2.63) .. controls (5.56,-1.12) and (2.65,-0.24) .. (0,0) .. controls (2.65,0.24) and (5.56,1.12) .. (8.74,2.63)   ;
\draw    (204.13,113.75) -- (260,170) ;
\draw [shift={(225,134.75)}, rotate = 45.41](8.74,-2.63) .. controls (5.56,-1.12) and (2.65,-0.24) .. (0,0) .. controls (2.65,0.24) and (5.56,1.12) .. (8.74,2.63)   ;
\draw [densely dashed]   (209.5,110) -- (259.5,160) ;
\draw[densely dashed]   (259.5,60) -- (209.13,110.5) ;
\draw [shift={(235.25,84.5)}, rotate = 134.63](8.74,-2.63) .. controls (5.56,-1.12) and (2.65,-0.24) .. (0,0) .. controls (2.65,0.24) and (5.56,1.12) .. (8.74,2.63)   ;
\draw [densely dashed]    (259.75,49.5) -- (203.88,106) ;
\draw [shift={(230,79.75)}, rotate = 134.58]  (8.74,-2.63) .. controls (5.56,-1.12) and (2.65,-0.24) .. (0,0) .. controls (2.65,0.24) and (5.56,1.12) .. (8.74,2.63)   ;
\draw [shift={(229.75,130)}, rotate = 45.41] (8.74,-2.63) .. controls (5.56,-1.12) and (2.65,-0.24) .. (0,0) .. controls (2.65,0.24) and (5.56,1.12) .. (8.74,2.63)   ;
\draw [densely dashed]    (115,106) .. controls (127.88,57.75) and (187.88,58.5) .. (203.88,106) ;
\draw    (115,114) .. controls (130.88,160.75) and (186.13,162.5) .. (203.88,114) ;
\draw    (119.88,110) .. controls (132.88,64) and (183.38,67.25) .. (197.88,110.25) ; 
\draw    (119.88,110) .. controls (136.38,153) and (180.38,155) .. (197.88,110.25) ;
\draw [shift={(160.5,70.25)}, rotate = 181.51] (8.74,-2.63) .. controls (5.56,-1.12) and (2.65,-0.24) .. (0,0) .. controls (2.65,0.24) and (5.56,1.12) .. (8.74,2.63)   ;
\draw [shift={(160.75,76.75)}, rotate = 181.51]  (8.74,-2.63) .. controls (5.56,-1.12) and (2.65,-0.24) .. (0,0) .. controls (2.65,0.24) and (5.56,1.12) .. (8.74,2.63)   ;
\draw [shift={(156,143)}, rotate = 355.91](8.74,-2.63) .. controls (5.56,-1.12) and (2.65,-0.24) .. (0,0) .. controls (2.65,0.24) and (5.56,1.12) .. (8.74,2.63)   ;
\draw [shift={(156.5,149.75)}, rotate = 355.91] (8.74,-2.63) .. controls (5.56,-1.12) and (2.65,-0.24) .. (0,0) .. controls (2.65,0.24) and (5.56,1.12) .. (8.74,2.63)   ;
\end{tikzpicture}}} = \dfrac{(q+1)}{2}g_5 g_7 I, \;\; 
\vcenter{\hbox{\begin{tikzpicture}[x=0.5pt,y=0.5pt,yscale=-1,xscale=1]
\draw [densely dashed]   (60,60) -- (109.88,110) ;
\draw [densely dashed]   (110.13,109.75) -- (60,160) ;
\draw[densely dashed]     (59.5,50) -- (98.75,89.75) -- (115,106) ;
\draw    (115,114) -- (59.5,170) ;
\draw  [shift={(95,85.75)}, rotate = 225] (8.74,-2.63) .. controls (5.56,-1.12) and (2.65,-0.24) .. (0,0) .. controls (2.65,0.24) and (5.56,1.12) .. (8.74,2.63)   ;
\draw  [shift={(90,90)}, rotate = 225]   (8.74,-2.63) .. controls (5.56,-1.12) and (2.65,-0.24) .. (0,0) .. controls (2.65,0.24) and (5.56,1.12) .. (8.74,2.63)   ;
\draw [shift={(80,140)}, rotate = 315]   (8.74,-2.63) .. controls (5.56,-1.12) and (2.65,-0.24) .. (0,0) .. controls (2.65,0.24) and (5.56,1.12) .. (8.74,2.63)   ;
\draw [shift={(84.75,144.5)}, rotate = 315]  (8.74,-2.63) .. controls (5.56,-1.12) and (2.65,-0.24) .. (0,0) .. controls (2.65,0.24) and (5.56,1.12) .. (8.74,2.63)   ;
\draw    (204.13,113.75) -- (260,170) ;
\draw [shift={(225,134.75)}, rotate = 45.41]    (8.74,-2.63) .. controls (5.56,-1.12) and (2.65,-0.24) .. (0,0) .. controls (2.65,0.24) and (5.56,1.12) .. (8.74,2.63)   ;
\draw [densely dashed]   (209.5,110) -- (259.5,160) ;
\draw[densely dashed]   (259.5,60) -- (209.13,110.5) ;
\draw [shift={(235.25,84.5)}, rotate = 134.63] (8.74,-2.63) .. controls (5.56,-1.12) and (2.65,-0.24) .. (0,0) .. controls (2.65,0.24) and (5.56,1.12) .. (8.74,2.63)   ;
\draw [densely dashed]    (259.75,49.5) -- (203.88,106) ;
\draw [shift={(230,79.75)}, rotate = 134.58](8.74,-2.63) .. controls (5.56,-1.12) and (2.65,-0.24) .. (0,0) .. controls (2.65,0.24) and (5.56,1.12) .. (8.74,2.63)   ;
\draw [shift={(229.75,130)}, rotate = 45.41]    (8.74,-2.63) .. controls (5.56,-1.12) and (2.65,-0.24) .. (0,0) .. controls (2.65,0.24) and (5.56,1.12) .. (8.74,2.63)   ;
\draw [densely dashed]    (115,106) .. controls (127.88,57.75) and (187.88,58.5) .. (203.88,106) ;
\draw    (115,114) .. controls (130.88,160.75) and (186.13,162.5) .. (203.88,114) ;
\draw [densely dashed]   (119.88,110) .. controls (132.88,64) and (183.38,67.25) .. (197.88,110.25) ; 
\draw [densely dashed]   (119.88,110) .. controls (136.38,153) and (180.38,155) .. (197.88,110.25) ;
\draw [shift={(160.5,70.25)}, rotate = 181.51](8.74,-2.63) .. controls (5.56,-1.12) and (2.65,-0.24) .. (0,0) .. controls (2.65,0.24) and (5.56,1.12) .. (8.74,2.63)   ;
\draw [shift={(160.75,76.75)}, rotate = 181.51]    (8.74,-2.63) .. controls (5.56,-1.12) and (2.65,-0.24) .. (0,0) .. controls (2.65,0.24) and (5.56,1.12) .. (8.74,2.63)   ;
\draw [shift={(156,143)}, rotate = 355.91](8.74,-2.63) .. controls (5.56,-1.12) and (2.65,-0.24) .. (0,0) .. controls (2.65,0.24) and (5.56,1.12) .. (8.74,2.63)   ;
\draw [shift={(156.5,149.75)}, rotate = 355.91](8.74,-2.63) .. controls (5.56,-1.12) and (2.65,-0.24) .. (0,0) .. controls (2.65,0.24) and (5.56,1.12) .. (8.74,2.63)   ;
\end{tikzpicture}}} = \dfrac{(p-1)}{2}g_6^2 I,\\
&\vcenter{\hbox{\begin{tikzpicture}[x=0.5pt,y=0.5pt,yscale=-1,xscale=1]
\draw [densely dashed]   (60,60) -- (109.88,110) ;
\draw [densely dashed]   (110.13,109.75) -- (60,160) ;
\draw[densely dashed]     (59.5,50) -- (98.75,89.75) -- (115,106) ;
\draw    (115,114) -- (59.5,170) ;
\draw  [shift={(95,85.75)}, rotate = 225] (8.74,-2.63) .. controls (5.56,-1.12) and (2.65,-0.24) .. (0,0) .. controls (2.65,0.24) and (5.56,1.12) .. (8.74,2.63)   ;
\draw  [shift={(90,90)}, rotate = 225]   (8.74,-2.63) .. controls (5.56,-1.12) and (2.65,-0.24) .. (0,0) .. controls (2.65,0.24) and (5.56,1.12) .. (8.74,2.63)   ;
\draw [shift={(80,140)}, rotate = 315]  (8.74,-2.63) .. controls (5.56,-1.12) and (2.65,-0.24) .. (0,0) .. controls (2.65,0.24) and (5.56,1.12) .. (8.74,2.63)   ;
\draw [shift={(84.75,144.5)}, rotate = 315](8.74,-2.63) .. controls (5.56,-1.12) and (2.65,-0.24) .. (0,0) .. controls (2.65,0.24) and (5.56,1.12) .. (8.74,2.63)   ;
\draw    (204.13,113.75) -- (260,170) ;
\draw [shift={(225,134.75)}, rotate = 45.41](8.74,-2.63) .. controls (5.56,-1.12) and (2.65,-0.24) .. (0,0) .. controls (2.65,0.24) and (5.56,1.12) .. (8.74,2.63)   ;
\draw   (209.5,110) -- (259.5,160) ;
\draw  (259.5,60) -- (209.13,110.5) ;
\draw [shift={(235.25,84.5)}, rotate = 134.63](8.74,-2.63) .. controls (5.56,-1.12) and (2.65,-0.24) .. (0,0) .. controls (2.65,0.24) and (5.56,1.12) .. (8.74,2.63)   ;
\draw [densely dashed]    (259.75,49.5) -- (203.88,106) ;
\draw [shift={(230,79.75)}, rotate = 134.58] (8.74,-2.63) .. controls (5.56,-1.12) and (2.65,-0.24) .. (0,0) .. controls (2.65,0.24) and (5.56,1.12) .. (8.74,2.63)   ;
\draw [shift={(229.75,130)}, rotate = 45.41] (8.74,-2.63) .. controls (5.56,-1.12) and (2.65,-0.24) .. (0,0) .. controls (2.65,0.24) and (5.56,1.12) .. (8.74,2.63)   ;
\draw [densely dashed]    (115,106) .. controls (127.88,57.75) and (187.88,58.5) .. (203.88,106) ;
\draw    (115,114) .. controls (130.88,160.75) and (186.13,162.5) .. (203.88,114) ;
\draw [densely dashed]   (119.88,110) .. controls (132.88,64) and (183.38,67.25) .. (197.88,110.25) ; 
\draw [densely dashed]   (119.88,110) .. controls (136.38,153) and (180.38,155) .. (197.88,110.25) ;
\draw [shift={(160.5,70.25)}, rotate = 181.51](8.74,-2.63) .. controls (5.56,-1.12) and (2.65,-0.24) .. (0,0) .. controls (2.65,0.24) and (5.56,1.12) .. (8.74,2.63)   ;
\draw [shift={(160.75,76.75)}, rotate = 181.51]   (8.74,-2.63) .. controls (5.56,-1.12) and (2.65,-0.24) .. (0,0) .. controls (2.65,0.24) and (5.56,1.12) .. (8.74,2.63)   ;
\draw [shift={(156,143)}, rotate = 355.91](8.74,-2.63) .. controls (5.56,-1.12) and (2.65,-0.24) .. (0,0) .. controls (2.65,0.24) and (5.56,1.12) .. (8.74,2.63)   ;
\draw [shift={(156.5,149.75)}, rotate = 355.91] (8.74,-2.63) .. controls (5.56,-1.12) and (2.65,-0.24) .. (0,0) .. controls (2.65,0.24) and (5.56,1.12) .. (8.74,2.63)   ;
\end{tikzpicture}}} = \dfrac{(p-1)}{2}g_6 g_7 I,\;\; 
\vcenter{\hbox{\begin{tikzpicture}[x=0.5pt,y=0.5pt,yscale=-1,xscale=1]
\draw  (60,60) -- (109.88,110) ;
\draw  (110.13,109.75) -- (60,160) ;
\draw[densely dashed]     (59.5,50) -- (98.75,89.75) -- (115,106) ;
\draw    (115,114) -- (59.5,170) ;
\draw  [shift={(95,85.75)}, rotate = 225] (8.74,-2.63) .. controls (5.56,-1.12) and (2.65,-0.24) .. (0,0) .. controls (2.65,0.24) and (5.56,1.12) .. (8.74,2.63)   ;
\draw  [shift={(90,90)}, rotate = 225]   (8.74,-2.63) .. controls (5.56,-1.12) and (2.65,-0.24) .. (0,0) .. controls (2.65,0.24) and (5.56,1.12) .. (8.74,2.63)   ;
\draw [shift={(80,140)}, rotate = 315]   (8.74,-2.63) .. controls (5.56,-1.12) and (2.65,-0.24) .. (0,0) .. controls (2.65,0.24) and (5.56,1.12) .. (8.74,2.63)   ;
\draw [shift={(84.75,144.5)}, rotate = 315]  (8.74,-2.63) .. controls (5.56,-1.12) and (2.65,-0.24) .. (0,0) .. controls (2.65,0.24) and (5.56,1.12) .. (8.74,2.63)   ;
\draw    (204.13,113.75) -- (260,170) ;
\draw [shift={(225,134.75)}, rotate = 45.41] (8.74,-2.63) .. controls (5.56,-1.12) and (2.65,-0.24) .. (0,0) .. controls (2.65,0.24) and (5.56,1.12) .. (8.74,2.63)   ;
\draw   (209.5,110) -- (259.5,160) ;
\draw  (259.5,60) -- (209.13,110.5) ;
\draw [shift={(235.25,84.5)}, rotate = 134.63](8.74,-2.63) .. controls (5.56,-1.12) and (2.65,-0.24) .. (0,0) .. controls (2.65,0.24) and (5.56,1.12) .. (8.74,2.63)   ;
\draw [densely dashed]    (259.75,49.5) -- (203.88,106) ;
\draw [shift={(230,79.75)}, rotate = 134.58] (8.74,-2.63) .. controls (5.56,-1.12) and (2.65,-0.24) .. (0,0) .. controls (2.65,0.24) and (5.56,1.12) .. (8.74,2.63)   ;
\draw [shift={(229.75,130)}, rotate = 45.41](8.74,-2.63) .. controls (5.56,-1.12) and (2.65,-0.24) .. (0,0) .. controls (2.65,0.24) and (5.56,1.12) .. (8.74,2.63)   ;
\draw [densely dashed]    (115,106) .. controls (127.88,57.75) and (187.88,58.5) .. (203.88,106) ;
\draw    (115,114) .. controls (130.88,160.75) and (186.13,162.5) .. (203.88,114) ;
\draw [densely dashed]   (119.88,110) .. controls (132.88,64) and (183.38,67.25) .. (197.88,110.25) ; 
\draw [densely dashed]   (119.88,110) .. controls (136.38,153) and (180.38,155) .. (197.88,110.25) ;
\draw [shift={(160.5,70.25)}, rotate = 181.51](8.74,-2.63) .. controls (5.56,-1.12) and (2.65,-0.24) .. (0,0) .. controls (2.65,0.24) and (5.56,1.12) .. (8.74,2.63)   ;
\draw [shift={(160.75,76.75)}, rotate = 181.51]   (8.74,-2.63) .. controls (5.56,-1.12) and (2.65,-0.24) .. (0,0) .. controls (2.65,0.24) and (5.56,1.12) .. (8.74,2.63)   ;
\draw [shift={(156,143)}, rotate = 355.91](8.74,-2.63) .. controls (5.56,-1.12) and (2.65,-0.24) .. (0,0) .. controls (2.65,0.24) and (5.56,1.12) .. (8.74,2.63)   ;
\draw [shift={(156.5,149.75)}, rotate = 355.91](8.74,-2.63) .. controls (5.56,-1.12) and (2.65,-0.24) .. (0,0) .. controls (2.65,0.24) and (5.56,1.12) .. (8.74,2.63)   ;
\end{tikzpicture}}} = \dfrac{(p-1)}{2}g_7^2 I,\\
&\vcenter{\hbox{\begin{tikzpicture}[x=0.5pt,y=0.5pt,yscale=-1,xscale=1]
\draw [densely dashed]   (60,60) -- (109.88,110) ;
\draw [densely dashed]   (110.13,109.75) -- (60,160) ;
\draw[densely dashed]     (59.5,50) -- (98.75,89.75) -- (115,106) ;
\draw    (115,114) -- (59.5,170) ;
\draw  [shift={(95,85.75)}, rotate = 225] (8.74,-2.63) .. controls (5.56,-1.12) and (2.65,-0.24) .. (0,0) .. controls (2.65,0.24) and (5.56,1.12) .. (8.74,2.63)   ;
\draw  [shift={(90,90)}, rotate = 225]  (8.74,-2.63) .. controls (5.56,-1.12) and (2.65,-0.24) .. (0,0) .. controls (2.65,0.24) and (5.56,1.12) .. (8.74,2.63)   ;
\draw [shift={(80,140)}, rotate = 315] (8.74,-2.63) .. controls (5.56,-1.12) and (2.65,-0.24) .. (0,0) .. controls (2.65,0.24) and (5.56,1.12) .. (8.74,2.63)   ;
\draw [shift={(84.75,144.5)}, rotate = 315] (8.74,-2.63) .. controls (5.56,-1.12) and (2.65,-0.24) .. (0,0) .. controls (2.65,0.24) and (5.56,1.12) .. (8.74,2.63)   ;
\draw    (204.13,113.75) -- (260,170) ;
\draw [shift={(225,134.75)}, rotate = 45.41] (8.74,-2.63) .. controls (5.56,-1.12) and (2.65,-0.24) .. (0,0) .. controls (2.65,0.24) and (5.56,1.12) .. (8.74,2.63)   ;
\draw [densely dashed]   (209.5,110) -- (259.5,160) ;
\draw [densely dashed]  (259.5,60) -- (209.13,110.5) ;
\draw [shift={(235.25,84.5)}, rotate = 134.63](8.74,-2.63) .. controls (5.56,-1.12) and (2.65,-0.24) .. (0,0) .. controls (2.65,0.24) and (5.56,1.12) .. (8.74,2.63)   ;
\draw [densely dashed]    (259.75,49.5) -- (203.88,106) ;
\draw [shift={(230,79.75)}, rotate = 134.58] (8.74,-2.63) .. controls (5.56,-1.12) and (2.65,-0.24) .. (0,0) .. controls (2.65,0.24) and (5.56,1.12) .. (8.74,2.63)   ;
\draw [shift={(229.75,130)}, rotate = 45.41] (8.74,-2.63) .. controls (5.56,-1.12) and (2.65,-0.24) .. (0,0) .. controls (2.65,0.24) and (5.56,1.12) .. (8.74,2.63)   ;
\draw [densely dashed]    (115,106) .. controls (127.88,57.75) and (187.88,58.5) .. (203.88,106) ;
\draw    (115,114) .. controls (130.88,160.75) and (186.13,162.5) .. (203.88,114) ;
\draw    (119.88,110) .. controls (132.88,64) and (183.38,67.25) .. (197.88,110.25) ; 
\draw   (119.88,110) .. controls (136.38,153) and (180.38,155) .. (197.88,110.25) ;
\draw [shift={(160.5,70.25)}, rotate = 181.51](8.74,-2.63) .. controls (5.56,-1.12) and (2.65,-0.24) .. (0,0) .. controls (2.65,0.24) and (5.56,1.12) .. (8.74,2.63)   ;
\draw [shift={(160.75,76.75)}, rotate = 181.51]  (8.74,-2.63) .. controls (5.56,-1.12) and (2.65,-0.24) .. (0,0) .. controls (2.65,0.24) and (5.56,1.12) .. (8.74,2.63)   ;
\draw [shift={(156,143)}, rotate = 355.91](8.74,-2.63) .. controls (5.56,-1.12) and (2.65,-0.24) .. (0,0) .. controls (2.65,0.24) and (5.56,1.12) .. (8.74,2.63)   ;
\draw [shift={(156.5,149.75)}, rotate = 355.91](8.74,-2.63) .. controls (5.56,-1.12) and (2.65,-0.24) .. (0,0) .. controls (2.65,0.24) and (5.56,1.12) .. (8.74,2.63)   ;
\end{tikzpicture}}} = \dfrac{(q+1)}{2}g_7^2 I,\;\;
\vcenter{\hbox{\begin{tikzpicture}[x=0.5pt,y=0.5pt,yscale=-1,xscale=1]
\draw   (60.25,55.5) -- (101.51,96.86) -- (110.13,105.5) ;
\draw     (110.13,115.5) -- (60,165.75) ;
\draw   [densely dashed]  (60.13,45.25) -- (93.75,79) -- (110,95.25) ;
\draw    (110,125.5) -- (60.13,175.75) ;
\draw [shift={(95.5,80.75)}, rotate = 225]   (8.74,-2.63) .. controls (5.56,-1.12) and (2.65,-0.24) .. (0,0) .. controls (2.65,0.24) and (5.56,1.12) .. (8.74,2.63)   ;
\draw [shift={(90.5,85.75)}, rotate = 225] (8.74,-2.63) .. controls (5.56,-1.12) and (2.65,-0.24) .. (0,0) .. controls (2.65,0.24) and (5.56,1.12) .. (8.74,2.63)   ;
\draw [shift={(77.75,148)}, rotate = 315] (8.74,-2.63) .. controls (5.56,-1.12) and (2.65,-0.24) .. (0,0) .. controls (2.65,0.24) and (5.56,1.12) .. (8.74,2.63)   ;
\draw [shift={(80.5,155.25)}, rotate = 315] (8.74,-2.63) .. controls (5.56,-1.12) and (2.65,-0.24) .. (0,0) .. controls (2.65,0.24) and (5.56,1.12) .. (8.74,2.63)   ;
\draw  (204.13,113.75) -- (260,170) ;
\draw [shift={(225,134.75)}, rotate = 45.41](8.74,-2.63) .. controls (5.56,-1.12) and (2.65,-0.24) .. (0,0) .. controls (2.65,0.24) and (5.56,1.12) .. (8.74,2.63)   ;
\draw   (209.5,110) -- (259.5,160) ;
\draw    (259.5,60) -- (209.13,110.5) ;
\draw [shift={(235.25,84.5)}, rotate = 134.63](8.74,-2.63) .. controls (5.56,-1.12) and (2.65,-0.24) .. (0,0) .. controls (2.65,0.24) and (5.56,1.12) .. (8.74,2.63)   ;
\draw  [densely dashed]   (259.75,49.5) -- (203.88,106) ;
\draw [shift={(230,79.75)}, rotate = 134.58] (8.74,-2.63) .. controls (5.56,-1.12) and (2.65,-0.24) .. (0,0) .. controls (2.65,0.24) and (5.56,1.12) .. (8.74,2.63)   ;
\draw [shift={(229.75,130)}, rotate = 45.41] (8.74,-2.63) .. controls (5.56,-1.12) and (2.65,-0.24) .. (0,0) .. controls (2.65,0.24) and (5.56,1.12) .. (8.74,2.63)   ;
\draw  [densely dashed]   (110,95.25) .. controls (124.13,62.5) and (184.13,58.75) .. (203.88,106) ;
\draw    (110,125.5) .. controls (127.88,159.25) and (186.38,162.25) .. (204.13,113.75) ;
\draw    (110.13,105.5) .. controls (129.13,66.75) and (183.38,67.25) .. (197.88,110.25) ;
\draw    (110.13,115.5) .. controls (130.13,153.5) and (180.38,155) .. (197.88,110.25) ;
\draw [shift={(157,70.5)}, rotate = 181.51](8.74,-2.63) .. controls (5.56,-1.12) and (2.65,-0.24) .. (0,0) .. controls (2.65,0.24) and (5.56,1.12) .. (8.74,2.63)   ;
\draw [shift={(157.25,77.25)}, rotate = 181.51] (8.74,-2.63) .. controls (5.56,-1.12) and (2.65,-0.24) .. (0,0) .. controls (2.65,0.24) and (5.56,1.12) .. (8.74,2.63)   ;
\draw [shift={(150.5,144)}, rotate = 355.91](8.74,-2.63) .. controls (5.56,-1.12) and (2.65,-0.24) .. (0,0) .. controls (2.65,0.24) and (5.56,1.12) .. (8.74,2.63)   ;
\draw [shift={(151,150.5)}, rotate = 355.91](8.74,-2.63) .. controls (5.56,-1.12) and (2.65,-0.24) .. (0,0) .. controls (2.65,0.24) and (5.56,1.12) .. (8.74,2.63)   ;
\end{tikzpicture}}} = \dfrac{1}{2}g_1g_5 I,\\
&\vcenter{\hbox{\begin{tikzpicture}[x=0.5pt,y=0.5pt,yscale=-1,xscale=1]
\draw   (60.25,55.5) -- (101.51,96.86) -- (110.13,105.5) ;
\draw    (110.13,115.5) -- (60,165.75) ;
\draw   [densely dashed]  (60.13,45.25) -- (93.75,79) -- (110,95.25) ;
\draw    (110,125.5) -- (60.13,175.75) ;
\draw [shift={(95.5,80.75)}, rotate = 225] (8.74,-2.63) .. controls (5.56,-1.12) and (2.65,-0.24) .. (0,0) .. controls (2.65,0.24) and (5.56,1.12) .. (8.74,2.63)   ;
\draw [shift={(90.5,85.75)}, rotate = 225] (8.74,-2.63) .. controls (5.56,-1.12) and (2.65,-0.24) .. (0,0) .. controls (2.65,0.24) and (5.56,1.12) .. (8.74,2.63)   ;
\draw [shift={(77.75,148)}, rotate = 315](8.74,-2.63) .. controls (5.56,-1.12) and (2.65,-0.24) .. (0,0) .. controls (2.65,0.24) and (5.56,1.12) .. (8.74,2.63)   ;
\draw [shift={(80.5,155.25)}, rotate = 315] (8.74,-2.63) .. controls (5.56,-1.12) and (2.65,-0.24) .. (0,0) .. controls (2.65,0.24) and (5.56,1.12) .. (8.74,2.63)   ;
\draw  (204.13,113.75) -- (260,170) ;
\draw [shift={(225,134.75)}, rotate = 45.41] (8.74,-2.63) .. controls (5.56,-1.12) and (2.65,-0.24) .. (0,0) .. controls (2.65,0.24) and (5.56,1.12) .. (8.74,2.63)   ;
\draw  [densely dashed] (209.5,110) -- (259.5,160) ;
\draw   [densely dashed] (259.5,60) -- (209.13,110.5) ;
\draw [shift={(235.25,84.5)}, rotate = 134.63](8.74,-2.63) .. controls (5.56,-1.12) and (2.65,-0.24) .. (0,0) .. controls (2.65,0.24) and (5.56,1.12) .. (8.74,2.63)   ;
\draw  [densely dashed]   (259.75,49.5) -- (203.88,106) ;
\draw [shift={(230,79.75)}, rotate = 134.58](8.74,-2.63) .. controls (5.56,-1.12) and (2.65,-0.24) .. (0,0) .. controls (2.65,0.24) and (5.56,1.12) .. (8.74,2.63)   ;
\draw [shift={(229.75,130)}, rotate = 45.41] (8.74,-2.63) .. controls (5.56,-1.12) and (2.65,-0.24) .. (0,0) .. controls (2.65,0.24) and (5.56,1.12) .. (8.74,2.63)   ;
\draw  [densely dashed]   (110,95.25) .. controls (124.13,62.5) and (184.13,58.75) .. (203.88,106) ;
\draw    (110,125.5) .. controls (127.88,159.25) and (186.38,162.25) .. (204.13,113.75) ;
\draw    (110.13,105.5) .. controls (129.13,66.75) and (183.38,67.25) .. (197.88,110.25) ;
\draw    (110.13,115.5) .. controls (130.13,153.5) and (180.38,155) .. (197.88,110.25) ;
\draw [shift={(157,70.5)}, rotate = 181.51](8.74,-2.63) .. controls (5.56,-1.12) and (2.65,-0.24) .. (0,0) .. controls (2.65,0.24) and (5.56,1.12) .. (8.74,2.63)   ;
\draw [shift={(157.25,77.25)}, rotate = 181.51] (8.74,-2.63) .. controls (5.56,-1.12) and (2.65,-0.24) .. (0,0) .. controls (2.65,0.24) and (5.56,1.12) .. (8.74,2.63)   ;
\draw [shift={(150.5,144)}, rotate = 355.91](8.74,-2.63) .. controls (5.56,-1.12) and (2.65,-0.24) .. (0,0) .. controls (2.65,0.24) and (5.56,1.12) .. (8.74,2.63)   ;
\draw [shift={(151,150.5)}, rotate = 355.91](8.74,-2.63) .. controls (5.56,-1.12) and (2.65,-0.24) .. (0,0) .. controls (2.65,0.24) and (5.56,1.12) .. (8.74,2.63)   ;
\end{tikzpicture}}} = \dfrac{1}{2}g_1g_7 I,\;\;\;\;\;\;\;\;
\vcenter{\hbox{\begin{tikzpicture}[x=0.5pt,y=0.5pt,yscale=-1,xscale=1]
\draw  [densely dashed]  (60.25,55.5) -- (101.51,96.86) -- (110.13,105.5) ;
\draw  [densely dashed]   (110.13,115.5) -- (60,165.75) ;
\draw   (60.13,45.25) -- (93.75,79) -- (110,95.25) ;
\draw  [densely dashed]   (110,125.5) -- (60.13,175.75) ;
\draw [shift={(95.5,80.75)}, rotate = 225] (8.74,-2.63) .. controls (5.56,-1.12) and (2.65,-0.24) .. (0,0) .. controls (2.65,0.24) and (5.56,1.12) .. (8.74,2.63)   ;
\draw [shift={(90.5,85.75)}, rotate = 225] (8.74,-2.63) .. controls (5.56,-1.12) and (2.65,-0.24) .. (0,0) .. controls (2.65,0.24) and (5.56,1.12) .. (8.74,2.63)   ;
\draw [shift={(77.75,148)}, rotate = 315] (8.74,-2.63) .. controls (5.56,-1.12) and (2.65,-0.24) .. (0,0) .. controls (2.65,0.24) and (5.56,1.12) .. (8.74,2.63)   ;
\draw [shift={(80.5,155.25)}, rotate = 315] (8.74,-2.63) .. controls (5.56,-1.12) and (2.65,-0.24) .. (0,0) .. controls (2.65,0.24) and (5.56,1.12) .. (8.74,2.63)   ;
\draw [densely dashed]  (204.13,113.75) -- (260,170) ;
\draw [shift={(225,134.75)}, rotate = 45.41]  (8.74,-2.63) .. controls (5.56,-1.12) and (2.65,-0.24) .. (0,0) .. controls (2.65,0.24) and (5.56,1.12) .. (8.74,2.63)   ;
\draw   (209.5,110) -- (259.5,160) ;
\draw    (259.5,60) -- (209.13,110.5) ;
\draw [shift={(235.25,84.5)}, rotate = 134.63](8.74,-2.63) .. controls (5.56,-1.12) and (2.65,-0.24) .. (0,0) .. controls (2.65,0.24) and (5.56,1.12) .. (8.74,2.63)   ;
\draw     (259.75,49.5) -- (203.88,106) ;
\draw [shift={(230,79.75)}, rotate = 134.58](8.74,-2.63) .. controls (5.56,-1.12) and (2.65,-0.24) .. (0,0) .. controls (2.65,0.24) and (5.56,1.12) .. (8.74,2.63)   ;
\draw [shift={(229.75,130)}, rotate = 45.41] (8.74,-2.63) .. controls (5.56,-1.12) and (2.65,-0.24) .. (0,0) .. controls (2.65,0.24) and (5.56,1.12) .. (8.74,2.63)   ;
\draw  (110,95.25) .. controls (124.13,62.5) and (184.13,58.75) .. (203.88,106) ;
\draw  [densely dashed]   (110,125.5) .. controls (127.88,159.25) and (186.38,162.25) .. (204.13,113.75) ;
\draw  [densely dashed]   (110.13,105.5) .. controls (129.13,66.75) and (183.38,67.25) .. (197.88,110.25) ;
\draw  [densely dashed]   (110.13,115.5) .. controls (130.13,153.5) and (180.38,155) .. (197.88,110.25) ;
\draw [shift={(157,70.5)}, rotate = 181.51](8.74,-2.63) .. controls (5.56,-1.12) and (2.65,-0.24) .. (0,0) .. controls (2.65,0.24) and (5.56,1.12) .. (8.74,2.63)   ;
\draw [shift={(157.25,77.25)}, rotate = 181.51] (8.74,-2.63) .. controls (5.56,-1.12) and (2.65,-0.24) .. (0,0) .. controls (2.65,0.24) and (5.56,1.12) .. (8.74,2.63)   ;
\draw [shift={(150.5,144)}, rotate = 355.91](8.74,-2.63) .. controls (5.56,-1.12) and (2.65,-0.24) .. (0,0) .. controls (2.65,0.24) and (5.56,1.12) .. (8.74,2.63)   ;
\draw [shift={(151,150.5)}, rotate = 355.91](8.74,-2.63) .. controls (5.56,-1.12) and (2.65,-0.24) .. (0,0) .. controls (2.65,0.24) and (5.56,1.12) .. (8.74,2.63)   ;
\end{tikzpicture}}} = \dfrac{1}{2}g_3g_7 I,\\ 
&\vcenter{\hbox{\begin{tikzpicture}[x=0.5pt,y=0.5pt,yscale=-1,xscale=1]
\draw  [densely dashed]  (60.25,55.5) -- (101.51,96.86) -- (110.13,105.5) ;
\draw  [densely dashed]   (110.13,115.5) -- (60,165.75) ;
\draw   (60.13,45.25) -- (93.75,79) -- (110,95.25) ;
\draw  [densely dashed]   (110,125.5) -- (60.13,175.75) ;
\draw [shift={(95.5,80.75)}, rotate = 225](8.74,-2.63) .. controls (5.56,-1.12) and (2.65,-0.24) .. (0,0) .. controls (2.65,0.24) and (5.56,1.12) .. (8.74,2.63)   ;
\draw [shift={(90.5,85.75)}, rotate = 225] (8.74,-2.63) .. controls (5.56,-1.12) and (2.65,-0.24) .. (0,0) .. controls (2.65,0.24) and (5.56,1.12) .. (8.74,2.63)   ;
\draw [shift={(77.75,148)}, rotate = 315](8.74,-2.63) .. controls (5.56,-1.12) and (2.65,-0.24) .. (0,0) .. controls (2.65,0.24) and (5.56,1.12) .. (8.74,2.63)   ;
\draw [shift={(80.5,155.25)}, rotate = 315] (8.74,-2.63) .. controls (5.56,-1.12) and (2.65,-0.24) .. (0,0) .. controls (2.65,0.24) and (5.56,1.12) .. (8.74,2.63)   ;
\draw [densely dashed]  (204.13,113.75) -- (260,170) ;
\draw [shift={(225,134.75)}, rotate = 45.41] (8.74,-2.63) .. controls (5.56,-1.12) and (2.65,-0.24) .. (0,0) .. controls (2.65,0.24) and (5.56,1.12) .. (8.74,2.63)   ;
\draw [densely dashed]  (209.5,110) -- (259.5,160) ;
\draw   [densely dashed] (259.5,60) -- (209.13,110.5) ;
\draw [shift={(235.25,84.5)}, rotate = 134.63](8.74,-2.63) .. controls (5.56,-1.12) and (2.65,-0.24) .. (0,0) .. controls (2.65,0.24) and (5.56,1.12) .. (8.74,2.63)   ;
\draw     (259.75,49.5) -- (203.88,106) ;
\draw [shift={(230,79.75)}, rotate = 134.58](8.74,-2.63) .. controls (5.56,-1.12) and (2.65,-0.24) .. (0,0) .. controls (2.65,0.24) and (5.56,1.12) .. (8.74,2.63)   ;
\draw [shift={(229.75,130)}, rotate = 45.41](8.74,-2.63) .. controls (5.56,-1.12) and (2.65,-0.24) .. (0,0) .. controls (2.65,0.24) and (5.56,1.12) .. (8.74,2.63)   ;
\draw  (110,95.25) .. controls (124.13,62.5) and (184.13,58.75) .. (203.88,106) ;
\draw  [densely dashed]   (110,125.5) .. controls (127.88,159.25) and (186.38,162.25) .. (204.13,113.75) ;
\draw  [densely dashed]   (110.13,105.5) .. controls (129.13,66.75) and (183.38,67.25) .. (197.88,110.25) ;
\draw  [densely dashed]   (110.13,115.5) .. controls (130.13,153.5) and (180.38,155) .. (197.88,110.25) ;
\draw [shift={(157,70.5)}, rotate = 181.51](8.74,-2.63) .. controls (5.56,-1.12) and (2.65,-0.24) .. (0,0) .. controls (2.65,0.24) and (5.56,1.12) .. (8.74,2.63)   ;
\draw [shift={(157.25,77.25)}, rotate = 181.51] (8.74,-2.63) .. controls (5.56,-1.12) and (2.65,-0.24) .. (0,0) .. controls (2.65,0.24) and (5.56,1.12) .. (8.74,2.63)   ;
\draw [shift={(150.5,144)}, rotate = 355.91] (8.74,-2.63) .. controls (5.56,-1.12) and (2.65,-0.24) .. (0,0) .. controls (2.65,0.24) and (5.56,1.12) .. (8.74,2.63)   ;
\draw [shift={(151,150.5)}, rotate = 355.91](8.74,-2.63) .. controls (5.56,-1.12) and (2.65,-0.24) .. (0,0) .. controls (2.65,0.24) and (5.56,1.12) .. (8.74,2.63)   ;
\end{tikzpicture}}} = \dfrac{1}{2}g_3g_6 I,\;\;\;\;\;\;\;\;
\vcenter{\hbox{\begin{tikzpicture}[x=0.5pt,y=0.5pt,yscale=-1,xscale=1]
\draw    (110,110) -- (59.88,160.25) ;
\draw   (113.33,115.33) -- (60.13,170.25) ;
\draw [shift={(90,90)}, rotate = 225]  (8.74,-2.63) .. controls (5.56,-1.12) and (2.65,-0.24) .. (0,0) .. controls (2.65,0.24) and (5.56,1.12) .. (8.74,2.63)   ;
\draw [shift={(79,141)}, rotate = 315]   (8.74,-2.63) .. controls (5.56,-1.12) and (2.65,-0.24) .. (0,0) .. controls (2.65,0.24) and (5.56,1.12) .. (8.74,2.63)   ;
\draw [shift={(80.67,149)}, rotate = 315]   (8.74,-2.63) .. controls (5.56,-1.12) and (2.65,-0.24) .. (0,0) .. controls (2.65,0.24) and (5.56,1.12) .. (8.74,2.63)   ;
\draw   (204.13,113.75) -- (260,170) ;
\draw [shift={(225,134.75)}, rotate = 45.41]    (8.74,-2.63) .. controls (5.56,-1.12) and (2.65,-0.24) .. (0,0) .. controls (2.65,0.24) and (5.56,1.12) .. (8.74,2.63)   ;
\draw   (209.5,110) -- (259.5,160) ;
\draw (259.5,60) -- (209.13,110.5) ;
\draw [shift={(235.25,84.5)}, rotate = 134.63]  (8.74,-2.63) .. controls (5.56,-1.12) and (2.65,-0.24) .. (0,0) .. controls (2.65,0.24) and (5.56,1.12) .. (8.74,2.63)   ;
\draw   [densely dashed]  (259.75,49.5) -- (203.88,106) ;
\draw [shift={(230,79.75)}, rotate = 134.58]    (8.74,-2.63) .. controls (5.56,-1.12) and (2.65,-0.24) .. (0,0) .. controls (2.65,0.24) and (5.56,1.12) .. (8.74,2.63)   ;
\draw [shift={(229.75,130)}, rotate = 45.41]  (8.74,-2.63) .. controls (5.56,-1.12) and (2.65,-0.24) .. (0,0) .. controls (2.65,0.24) and (5.56,1.12) .. (8.74,2.63)   ;
\draw  [densely dashed]   (114.33,103) .. controls (126.67,58.67) and (186.67,58) .. (203.88,106) ;
\draw    (120,110) .. controls (129.67,63.67) and (184.67,65) .. (197.88,110.25) ;
\draw [shift={(159,69.83)}, rotate = 181.51]    (8.74,-2.63) .. controls (5.56,-1.12) and (2.65,-0.24) .. (0,0) .. controls (2.65,0.24) and (5.56,1.12) .. (8.74,2.63)   ;
\draw [shift={(159.58,75.92)}, rotate = 181.51]   (8.74,-2.63) .. controls (5.56,-1.12) and (2.65,-0.24) .. (0,0) .. controls (2.65,0.24) and (5.56,1.12) .. (8.74,2.63)   ;
\draw  (60,60) -- (110,110) ;
\draw  [densely dashed]   (60,50) -- (114.33,103) ;
\draw [shift={(92.17,81.5)}, rotate = 225]   (8.74,-2.63) .. controls (5.56,-1.12) and (2.65,-0.24) .. (0,0) .. controls (2.65,0.24) and (5.56,1.12) .. (8.74,2.63)   ;
\draw    (113.33,115.33) .. controls (116,131.75) and (134.5,152) .. (146.5,150) .. controls (158.5,148) and (157.67,143.83) .. (167.75,142.75) .. controls (177.83,141.67) and (196.75,127.75) .. (197.88,110.25) ; 
\draw    (120,110) .. controls (119.75,125.25) and (130.5,138.25) .. (142,142.75) ;
\draw    (142,142.75) .. controls (148.75,146.5) and (150.5,145.25) .. (152.75,146.25) ;
\draw    (169,150.25) .. controls (184,144.75) and (197.5,136.75) .. (204.13,113.75) ;
\draw    (159,148) .. controls (162.75,148.5) and (164,151.5) .. (169,150.25) ;
\draw [shift={(172,142.08)}, rotate = 334.62]  (8.74,-2.63) .. controls (5.56,-1.12) and (2.65,-0.24) .. (0,0) .. controls (2.65,0.24) and (5.56,1.12) .. (8.74,2.63)   ;
\draw [shift={(134.5,138.58)}, rotate = 30.72]  (8.74,-2.63) .. controls (5.56,-1.12) and (2.65,-0.24) .. (0,0) .. controls (2.65,0.24) and (5.56,1.12) .. (8.74,2.63)   ;
\end{tikzpicture}}}=\dfrac{1}{2}g_5^2 I,\\
&\vcenter{\hbox{\begin{tikzpicture}[x=0.5pt,y=0.5pt,yscale=-1,xscale=1]
\draw    (110,110) -- (59.88,160.25) ;
\draw   (113.33,115.33) -- (60.13,170.25) ;
\draw [shift={(90,90)}, rotate = 225]  (8.74,-2.63) .. controls (5.56,-1.12) and (2.65,-0.24) .. (0,0) .. controls (2.65,0.24) and (5.56,1.12) .. (8.74,2.63)   ;
\draw [shift={(79,141)}, rotate = 315]   (8.74,-2.63) .. controls (5.56,-1.12) and (2.65,-0.24) .. (0,0) .. controls (2.65,0.24) and (5.56,1.12) .. (8.74,2.63)   ;
\draw [shift={(80.67,149)}, rotate = 315]   (8.74,-2.63) .. controls (5.56,-1.12) and (2.65,-0.24) .. (0,0) .. controls (2.65,0.24) and (5.56,1.12) .. (8.74,2.63)   ;
\draw (204.13,113.75) -- (260,170) ;
\draw [shift={(225,134.75)}, rotate = 45.41]    (8.74,-2.63) .. controls (5.56,-1.12) and (2.65,-0.24) .. (0,0) .. controls (2.65,0.24) and (5.56,1.12) .. (8.74,2.63)   ;
\draw  [densely dashed]  (209.5,110) -- (259.5,160) ;
\draw [densely dashed] (259.5,60) -- (209.13,110.5) ;
\draw [shift={(235.25,84.5)}, rotate = 134.63]  (8.74,-2.63) .. controls (5.56,-1.12) and (2.65,-0.24) .. (0,0) .. controls (2.65,0.24) and (5.56,1.12) .. (8.74,2.63)   ;
\draw   [densely dashed]  (259.75,49.5) -- (203.88,106) ;
\draw [shift={(230,79.75)}, rotate = 134.58]    (8.74,-2.63) .. controls (5.56,-1.12) and (2.65,-0.24) .. (0,0) .. controls (2.65,0.24) and (5.56,1.12) .. (8.74,2.63)   ;
\draw [shift={(229.75,130)}, rotate = 45.41]  (8.74,-2.63) .. controls (5.56,-1.12) and (2.65,-0.24) .. (0,0) .. controls (2.65,0.24) and (5.56,1.12) .. (8.74,2.63)   ;
\draw  [densely dashed]   (114.33,103) .. controls (126.67,58.67) and (186.67,58) .. (203.88,106) ;
\draw    (120,110) .. controls (129.67,63.67) and (184.67,65) .. (197.88,110.25) ;
\draw [shift={(159,69.83)}, rotate = 181.51]    (8.74,-2.63) .. controls (5.56,-1.12) and (2.65,-0.24) .. (0,0) .. controls (2.65,0.24) and (5.56,1.12) .. (8.74,2.63)   ;
\draw [shift={(159.58,75.92)}, rotate = 181.51]   (8.74,-2.63) .. controls (5.56,-1.12) and (2.65,-0.24) .. (0,0) .. controls (2.65,0.24) and (5.56,1.12) .. (8.74,2.63)   ;
\draw  (60,60) -- (110,110) ;
\draw  [densely dashed]   (60,50) -- (114.33,103) ;
\draw [shift={(92.17,81.5)}, rotate = 225]   (8.74,-2.63) .. controls (5.56,-1.12) and (2.65,-0.24) .. (0,0) .. controls (2.65,0.24) and (5.56,1.12) .. (8.74,2.63)   ;
\draw    (113.33,115.33) .. controls (116,131.75) and (134.5,152) .. (146.5,150) .. controls (158.5,148) and (157.67,143.83) .. (167.75,142.75) .. controls (177.83,141.67) and (196.75,127.75) .. (197.88,110.25) ; 
\draw    (120,110) .. controls (119.75,125.25) and (130.5,138.25) .. (142,142.75) ;
\draw    (142,142.75) .. controls (148.75,146.5) and (150.5,145.25) .. (152.75,146.25) ;
\draw    (169,150.25) .. controls (184,144.75) and (197.5,136.75) .. (204.13,113.75) ;
\draw    (159,148) .. controls (162.75,148.5) and (164,151.5) .. (169,150.25) ;
\draw [shift={(172,142.08)}, rotate = 334.62]  (8.74,-2.63) .. controls (5.56,-1.12) and (2.65,-0.24) .. (0,0) .. controls (2.65,0.24) and (5.56,1.12) .. (8.74,2.63)   ;
\draw [shift={(134.5,138.58)}, rotate = 30.72]  (8.74,-2.63) .. controls (5.56,-1.12) and (2.65,-0.24) .. (0,0) .. controls (2.65,0.24) and (5.56,1.12) .. (8.74,2.63)   ;
\end{tikzpicture}}}=\dfrac{1}{2}g_5g_7 I,\;\;\;\;\;\;\;\;
\vcenter{\hbox{\begin{tikzpicture}[x=0.5pt,y=0.5pt,yscale=-1,xscale=1]
\draw  [densely dashed]  (110,110) -- (59.88,160.25) ;
\draw   (113.33,115.33) -- (60.13,170.25) ;
\draw [shift={(90,90)}, rotate = 225]  (8.74,-2.63) .. controls (5.56,-1.12) and (2.65,-0.24) .. (0,0) .. controls (2.65,0.24) and (5.56,1.12) .. (8.74,2.63)   ;
\draw [shift={(79,141)}, rotate = 315]   (8.74,-2.63) .. controls (5.56,-1.12) and (2.65,-0.24) .. (0,0) .. controls (2.65,0.24) and (5.56,1.12) .. (8.74,2.63)   ;
\draw [shift={(80.67,149)}, rotate = 315]   (8.74,-2.63) .. controls (5.56,-1.12) and (2.65,-0.24) .. (0,0) .. controls (2.65,0.24) and (5.56,1.12) .. (8.74,2.63)   ;
\draw (204.13,113.75) -- (260,170) ;
\draw [shift={(225,134.75)}, rotate = 45.41]    (8.74,-2.63) .. controls (5.56,-1.12) and (2.65,-0.24) .. (0,0) .. controls (2.65,0.24) and (5.56,1.12) .. (8.74,2.63)   ;
\draw  [densely dashed]  (209.5,110) -- (259.5,160) ;
\draw [densely dashed] (259.5,60) -- (209.13,110.5) ;
\draw [shift={(235.25,84.5)}, rotate = 134.63]  (8.74,-2.63) .. controls (5.56,-1.12) and (2.65,-0.24) .. (0,0) .. controls (2.65,0.24) and (5.56,1.12) .. (8.74,2.63)   ;
\draw   [densely dashed]  (259.75,49.5) -- (203.88,106) ;
\draw [shift={(230,79.75)}, rotate = 134.58]    (8.74,-2.63) .. controls (5.56,-1.12) and (2.65,-0.24) .. (0,0) .. controls (2.65,0.24) and (5.56,1.12) .. (8.74,2.63)   ;
\draw [shift={(229.75,130)}, rotate = 45.41]  (8.74,-2.63) .. controls (5.56,-1.12) and (2.65,-0.24) .. (0,0) .. controls (2.65,0.24) and (5.56,1.12) .. (8.74,2.63)   ;
\draw  [densely dashed]   (114.33,103) .. controls (126.67,58.67) and (186.67,58) .. (203.88,106) ;
\draw    (120,110) .. controls (129.67,63.67) and (184.67,65) .. (197.88,110.25) ;
\draw [shift={(159,69.83)}, rotate = 181.51]    (8.74,-2.63) .. controls (5.56,-1.12) and (2.65,-0.24) .. (0,0) .. controls (2.65,0.24) and (5.56,1.12) .. (8.74,2.63)   ;
\draw [shift={(159.58,75.92)}, rotate = 181.51]   (8.74,-2.63) .. controls (5.56,-1.12) and (2.65,-0.24) .. (0,0) .. controls (2.65,0.24) and (5.56,1.12) .. (8.74,2.63)   ;
\draw  [densely dashed] (60,60) -- (110,110) ;
\draw  [densely dashed]   (60,50) -- (114.33,103) ;
\draw [shift={(92.17,81.5)}, rotate = 225]   (8.74,-2.63) .. controls (5.56,-1.12) and (2.65,-0.24) .. (0,0) .. controls (2.65,0.24) and (5.56,1.12) .. (8.74,2.63)   ;
\draw    (113.33,115.33) .. controls (116,131.75) and (134.5,152) .. (146.5,150) .. controls (158.5,148) and (157.67,143.83) .. (167.75,142.75) .. controls (177.83,141.67) and (196.75,127.75) .. (197.88,110.25) ; 
\draw    (120,110) .. controls (119.75,125.25) and (130.5,138.25) .. (142,142.75) ;
\draw    (142,142.75) .. controls (148.75,146.5) and (150.5,145.25) .. (152.75,146.25) ;
\draw    (169,150.25) .. controls (184,144.75) and (197.5,136.75) .. (204.13,113.75) ;
\draw    (159,148) .. controls (162.75,148.5) and (164,151.5) .. (169,150.25) ;
\draw [shift={(172,142.08)}, rotate = 334.62]  (8.74,-2.63) .. controls (5.56,-1.12) and (2.65,-0.24) .. (0,0) .. controls (2.65,0.24) and (5.56,1.12) .. (8.74,2.63)   ;
\draw [shift={(134.5,138.58)}, rotate = 30.72]  (8.74,-2.63) .. controls (5.56,-1.12) and (2.65,-0.24) .. (0,0) .. controls (2.65,0.24) and (5.56,1.12) .. (8.74,2.63)   ;
\end{tikzpicture}}}=\dfrac{1}{2}g_7^2 I,\\
&\vcenter{\hbox{\begin{tikzpicture}[x=0.5pt,y=0.5pt,yscale=-1,xscale=1]
\draw   (110,110) -- (59.88,160.25) ;
\draw    (113.25,117) -- (60.13,170.25) ;
\draw [shift={(90,90)}, rotate = 225] (8.74,-2.63) .. controls (5.56,-1.12) and (2.65,-0.24) .. (0,0) .. controls (2.65,0.24) and (5.56,1.12) .. (8.74,2.63)   ;
\draw [shift={(79,141)}, rotate = 315] (8.74,-2.63) .. controls (5.56,-1.12) and (2.65,-0.24) .. (0,0) .. controls (2.65,0.24) and (5.56,1.12) .. (8.74,2.63)   ;
\draw [shift={(81.17,149.25)}, rotate = 315]  (8.74,-2.63) .. controls (5.56,-1.12) and (2.65,-0.24) .. (0,0) .. controls (2.65,0.24) and (5.56,1.12) .. (8.74,2.63)   ;
\draw    (204.13,113.75) -- (260,170) ;
\draw [shift={(225,134.75)}, rotate = 45.41]   (8.74,-2.63) .. controls (5.56,-1.12) and (2.65,-0.24) .. (0,0) .. controls (2.65,0.24) and (5.56,1.12) .. (8.74,2.63)   ;
\draw     (209.5,110) -- (259.5,160) ;
\draw     (259.5,60) -- (209.13,110.5) ;
\draw [shift={(235.25,84.5)}, rotate = 134.63]   (8.74,-2.63) .. controls (5.56,-1.12) and (2.65,-0.24) .. (0,0) .. controls (2.65,0.24) and (5.56,1.12) .. (8.74,2.63)   ;
\draw  [densely dashed]   (259.75,49.5) -- (203.88,106) ;
\draw [shift={(230,79.75)}, rotate = 134.58]  (8.74,-2.63) .. controls (5.56,-1.12) and (2.65,-0.24) .. (0,0) .. controls (2.65,0.24) and (5.56,1.12) .. (8.74,2.63)   ;
\draw [shift={(229.75,130)}, rotate = 45.41]  (8.74,-2.63) .. controls (5.56,-1.12) and (2.65,-0.24) .. (0,0) .. controls (2.65,0.24) and (5.56,1.12) .. (8.74,2.63)   ;
\draw [shift={(155.25,143.83)}, rotate = 359.17]  (8.74,-2.63) .. controls (5.56,-1.12) and (2.65,-0.24) .. (0,0) .. controls (2.65,0.24) and (5.56,1.12) .. (8.74,2.63)   ;
\draw [shift={(156.08,150.67)}, rotate = 358.39] (8.74,-2.63) .. controls (5.56,-1.12) and (2.65,-0.24) .. (0,0) .. controls (2.65,0.24) and (5.56,1.12) .. (8.74,2.63)   ;
\draw   (60,60) -- (110,110) ;
\draw  [densely dashed]  (60,50) -- (114.33,103) ;
\draw [shift={(92.17,81.5)}, rotate = 225] (8.74,-2.63) .. controls (5.56,-1.12) and (2.65,-0.24) .. (0,0) .. controls (2.65,0.24) and (5.56,1.12) .. (8.74,2.63)   ;
\draw   (113.25,117) .. controls (129.75,162.25) and (186.5,162.25) .. (204.13,113.75) ;
\draw   [densely dashed]  (118.25,110) .. controls (131,156.25) and (185.75,154) .. (197.88,110.25) ;
\draw  [densely dashed]   (114.33,103) .. controls (117.5,89) and (130.5,72.25) .. (144.25,70.5) .. controls (158,68.75) and (153.75,76) .. (162.5,76.25) .. controls (171.25,76.5) and (191,85.75) .. (197.88,110.25) ;
\draw  [densely dashed]   (118.25,110) .. controls (121,101.25) and (130.75,78) .. (148.25,77.75) ;
\draw  [densely dashed]   (148.25,77.75) .. controls (150.75,77.75) and (152.75,75.75) .. (154.5,75.25) ;
\draw [densely dashed]    (162.25,70.5) .. controls (187,69.75) and (200.25,96.75) .. (203.88,106) ;
\draw  [densely dashed]   (157.75,72.75) .. controls (160.75,71.75) and (160.25,70.75) .. (162.25,70.5) ;
\draw [shift={(143.17,78.5)}, rotate = 157.75]  (8.74,-2.63) .. controls (5.56,-1.12) and (2.65,-0.24) .. (0,0) .. controls (2.65,0.24) and (5.56,1.12) .. (8.74,2.63)   ;
\draw [shift={(176.42,81)}, rotate = 205.51]  (8.74,-2.63) .. controls (5.56,-1.12) and (2.65,-0.24) .. (0,0) .. controls (2.65,0.24) and (5.56,1.12) .. (8.74,2.63) ;
\end{tikzpicture}}}=\dfrac{1}{2}g_7^2 I,\;\;\;\;\;\;\;\;\;\;\;
\vcenter{\hbox{\begin{tikzpicture}[x=0.5pt,y=0.5pt,yscale=-1,xscale=1]
\draw  [densely dashed] (110,110) -- (59.88,160.25) ;
\draw    (113.25,117) -- (60.13,170.25) ;
\draw [shift={(90,90)}, rotate = 225] (8.74,-2.63) .. controls (5.56,-1.12) and (2.65,-0.24) .. (0,0) .. controls (2.65,0.24) and (5.56,1.12) .. (8.74,2.63)   ;
\draw [shift={(79,141)}, rotate = 315] (8.74,-2.63) .. controls (5.56,-1.12) and (2.65,-0.24) .. (0,0) .. controls (2.65,0.24) and (5.56,1.12) .. (8.74,2.63)   ;
\draw [shift={(81.17,149.25)}, rotate = 315]  (8.74,-2.63) .. controls (5.56,-1.12) and (2.65,-0.24) .. (0,0) .. controls (2.65,0.24) and (5.56,1.12) .. (8.74,2.63)   ;
\draw    (204.13,113.75) -- (260,170) ;
\draw [shift={(225,134.75)}, rotate = 45.41]   (8.74,-2.63) .. controls (5.56,-1.12) and (2.65,-0.24) .. (0,0) .. controls (2.65,0.24) and (5.56,1.12) .. (8.74,2.63)   ;
\draw     (209.5,110) -- (259.5,160) ;
\draw     (259.5,60) -- (209.13,110.5) ;
\draw [shift={(235.25,84.5)}, rotate = 134.63]   (8.74,-2.63) .. controls (5.56,-1.12) and (2.65,-0.24) .. (0,0) .. controls (2.65,0.24) and (5.56,1.12) .. (8.74,2.63)   ;
\draw  [densely dashed]   (259.75,49.5) -- (203.88,106) ;
\draw [shift={(230,79.75)}, rotate = 134.58]  (8.74,-2.63) .. controls (5.56,-1.12) and (2.65,-0.24) .. (0,0) .. controls (2.65,0.24) and (5.56,1.12) .. (8.74,2.63)   ;
\draw [shift={(229.75,130)}, rotate = 45.41]  (8.74,-2.63) .. controls (5.56,-1.12) and (2.65,-0.24) .. (0,0) .. controls (2.65,0.24) and (5.56,1.12) .. (8.74,2.63)   ;
\draw [shift={(155.25,143.83)}, rotate = 359.17]  (8.74,-2.63) .. controls (5.56,-1.12) and (2.65,-0.24) .. (0,0) .. controls (2.65,0.24) and (5.56,1.12) .. (8.74,2.63)   ;
\draw [shift={(156.08,150.67)}, rotate = 358.39] (8.74,-2.63) .. controls (5.56,-1.12) and (2.65,-0.24) .. (0,0) .. controls (2.65,0.24) and (5.56,1.12) .. (8.74,2.63)   ;
\draw   [densely dashed] (60,60) -- (110,110) ;
\draw  [densely dashed]  (60,50) -- (114.33,103) ;
\draw [shift={(92.17,81.5)}, rotate = 225] (8.74,-2.63) .. controls (5.56,-1.12) and (2.65,-0.24) .. (0,0) .. controls (2.65,0.24) and (5.56,1.12) .. (8.74,2.63)   ;
\draw   (113.25,117) .. controls (129.75,162.25) and (186.5,162.25) .. (204.13,113.75) ;
\draw   [densely dashed]  (118.25,110) .. controls (131,156.25) and (185.75,154) .. (197.88,110.25) ;
\draw  [densely dashed]   (114.33,103) .. controls (117.5,89) and (130.5,72.25) .. (144.25,70.5) .. controls (158,68.75) and (153.75,76) .. (162.5,76.25) .. controls (171.25,76.5) and (191,85.75) .. (197.88,110.25) ;
\draw  [densely dashed]   (118.25,110) .. controls (121,101.25) and (130.75,78) .. (148.25,77.75) ;
\draw  [densely dashed]   (148.25,77.75) .. controls (150.75,77.75) and (152.75,75.75) .. (154.5,75.25) ;
\draw [densely dashed]    (162.25,70.5) .. controls (187,69.75) and (200.25,96.75) .. (203.88,106) ;
\draw  [densely dashed]   (157.75,72.75) .. controls (160.75,71.75) and (160.25,70.75) .. (162.25,70.5) ;
\draw [shift={(143.17,78.5)}, rotate = 157.75]  (8.74,-2.63) .. controls (5.56,-1.12) and (2.65,-0.24) .. (0,0) .. controls (2.65,0.24) and (5.56,1.12) .. (8.74,2.63)   ;
\draw [shift={(176.42,81)}, rotate = 205.51]  (8.74,-2.63) .. controls (5.56,-1.12) and (2.65,-0.24) .. (0,0) .. controls (2.65,0.24) and (5.56,1.12) .. (8.74,2.63) ;
\end{tikzpicture}}}=\dfrac{1}{2}g_6g_7 I,\\
&\vcenter{\hbox{\begin{tikzpicture}[x=0.5pt,y=0.5pt,yscale=-1,xscale=1]
\draw  [densely dashed] (110,110) -- (59.88,160.25) ;
\draw    (113.25,117) -- (60.13,170.25) ;
\draw [shift={(90,90)}, rotate = 225] (8.74,-2.63) .. controls (5.56,-1.12) and (2.65,-0.24) .. (0,0) .. controls (2.65,0.24) and (5.56,1.12) .. (8.74,2.63)   ;
\draw [shift={(79,141)}, rotate = 315] (8.74,-2.63) .. controls (5.56,-1.12) and (2.65,-0.24) .. (0,0) .. controls (2.65,0.24) and (5.56,1.12) .. (8.74,2.63)   ;
\draw [shift={(81.17,149.25)}, rotate = 315]  (8.74,-2.63) .. controls (5.56,-1.12) and (2.65,-0.24) .. (0,0) .. controls (2.65,0.24) and (5.56,1.12) .. (8.74,2.63)   ;
\draw    (204.13,113.75) -- (260,170) ;
\draw [shift={(225,134.75)}, rotate = 45.41]   (8.74,-2.63) .. controls (5.56,-1.12) and (2.65,-0.24) .. (0,0) .. controls (2.65,0.24) and (5.56,1.12) .. (8.74,2.63)   ;
\draw  [densely dashed]   (209.5,110) -- (259.5,160) ;
\draw  [densely dashed]   (259.5,60) -- (209.13,110.5) ;
\draw [shift={(235.25,84.5)}, rotate = 134.63]   (8.74,-2.63) .. controls (5.56,-1.12) and (2.65,-0.24) .. (0,0) .. controls (2.65,0.24) and (5.56,1.12) .. (8.74,2.63)   ;
\draw  [densely dashed]   (259.75,49.5) -- (203.88,106) ;
\draw [shift={(230,79.75)}, rotate = 134.58]  (8.74,-2.63) .. controls (5.56,-1.12) and (2.65,-0.24) .. (0,0) .. controls (2.65,0.24) and (5.56,1.12) .. (8.74,2.63)   ;
\draw [shift={(229.75,130)}, rotate = 45.41]  (8.74,-2.63) .. controls (5.56,-1.12) and (2.65,-0.24) .. (0,0) .. controls (2.65,0.24) and (5.56,1.12) .. (8.74,2.63)   ;
\draw [shift={(155.25,143.83)}, rotate = 359.17]  (8.74,-2.63) .. controls (5.56,-1.12) and (2.65,-0.24) .. (0,0) .. controls (2.65,0.24) and (5.56,1.12) .. (8.74,2.63)   ;
\draw [shift={(156.08,150.67)}, rotate = 358.39] (8.74,-2.63) .. controls (5.56,-1.12) and (2.65,-0.24) .. (0,0) .. controls (2.65,0.24) and (5.56,1.12) .. (8.74,2.63)   ;
\draw   [densely dashed] (60,60) -- (110,110) ;
\draw  [densely dashed]  (60,50) -- (114.33,103) ;
\draw [shift={(92.17,81.5)}, rotate = 225] (8.74,-2.63) .. controls (5.56,-1.12) and (2.65,-0.24) .. (0,0) .. controls (2.65,0.24) and (5.56,1.12) .. (8.74,2.63)   ;
\draw   (113.25,117) .. controls (129.75,162.25) and (186.5,162.25) .. (204.13,113.75) ;
\draw   [densely dashed]  (118.25,110) .. controls (131,156.25) and (185.75,154) .. (197.88,110.25) ;
\draw  [densely dashed]   (114.33,103) .. controls (117.5,89) and (130.5,72.25) .. (144.25,70.5) .. controls (158,68.75) and (153.75,76) .. (162.5,76.25) .. controls (171.25,76.5) and (191,85.75) .. (197.88,110.25) ;
\draw  [densely dashed]   (118.25,110) .. controls (121,101.25) and (130.75,78) .. (148.25,77.75) ;
\draw  [densely dashed]   (148.25,77.75) .. controls (150.75,77.75) and (152.75,75.75) .. (154.5,75.25) ;
\draw [densely dashed]    (162.25,70.5) .. controls (187,69.75) and (200.25,96.75) .. (203.88,106) ;
\draw  [densely dashed]   (157.75,72.75) .. controls (160.75,71.75) and (160.25,70.75) .. (162.25,70.5) ;
\draw [shift={(143.17,78.5)}, rotate = 157.75]  (8.74,-2.63) .. controls (5.56,-1.12) and (2.65,-0.24) .. (0,0) .. controls (2.65,0.24) and (5.56,1.12) .. (8.74,2.63)   ;
\draw [shift={(176.42,81)}, rotate = 205.51]  (8.74,-2.63) .. controls (5.56,-1.12) and (2.65,-0.24) .. (0,0) .. controls (2.65,0.24) and (5.56,1.12) .. (8.74,2.63) ;
\end{tikzpicture}}}=\dfrac{1}{2}g_6^2 I.
\end{align*}
\nn
\endgroup
Only the first six diagrams scale with $q$ or $p$ due to the closed flavour loops. The remaining diagrams have no such loops because of the vertices or because they are non-planar. Summing these and using the renormalisation scheme discussed above in the chiral Gross-Neveu model results in the beta-functions
\be\begin{aligned}
\beta(g_5)&=-\dfrac{1}{2\pi}\left[g_1g_5+g_5^2(N+p)+g_7^2p\right]\\
\beta(g_6)&=-\dfrac{1}{2\pi}\left[g_3g_6+g_6^2p+g_7^2(N+p)\right]\\
\beta(g_7)&=-\dfrac{1}{2\pi}\left[g_1 g_7+g_3g_7+g_5g_7(N+p)+g_6g_7p \right]
\end{aligned}\nn\ee

\acknowledgments
I would like to thank David Tong for suggesting this project, for many insightful discussions and for detailed comments on earlier drafts of this paper. I am very grateful for his PhD supervision. I am also very thankful to Avner Karasik and Nathan McStay for many enlightening conversations and comments on earlier drafts. I would further like to thank Alejandra Castro, Diego Delmastro, Nick Dorey, Jackson Fliss, Jaume Gomis and Christopher Hughes for discussions and correspondence on various topics. I am supported by a Cambridge Trust International Scholarship, the STFC consolidated grant ST/P000681/1 and the EPSRC grant EP/V047655/1 ``Chiral Gauge Theories: From Strong Coupling to the Standard Model". For the purpose of open access, I have applied a Creative Commons Attribution (CC BY) licence to any Author Accepted Manuscript version arising from this submission.

\end{document}